\documentclass[aps,pra,twocolumn,superscriptaddress,showpacs, amsmath, amssymb]{revtex4}
\usepackage[dvipdfmx]{graphicx}
\usepackage[dvipdfmx]{graphics}
\usepackage{bm}% bold math
\usepackage{amsfonts}
\usepackage{latexsym}

\usepackage[usenames]{color}
\usepackage{ulem}
\begin{document}
%%%%%%%%%%%%%%%%%%%%%%%%%%%%%%%%%%%%%%%%%%%%%%%%%%%%%%%
%Title of paper
\title{Theoretical Analysis on Spectroscopy of Atomic Bose-Hubbard Systems}
%Authors
\author{Kensuke Inaba}
\affiliation{NTT Basic Research Laboratories, NTT Corporation, Atsugi 243-0198, Japan}
\affiliation{JST, CREST, Chiyoda-ku, Tokyo 102-0075, Japan}

\author{Makoto Yamashita}
\affiliation{NTT Basic Research Laboratories, NTT Corporation, Atsugi 243-0198, Japan}
\affiliation{JST, CREST, Chiyoda-ku, Tokyo 102-0075, Japan}

\date{\today}

\begin{abstract}
We provide a numerical method to calculate comprehensively the microwave and the laser spectra of ultracold bosonic atoms in optical lattices at finite temperatures.
Our formulation is built up with the sum rules, up to the second order, derived from the general principle of spectroscopy.
The sum rule approach allows us to discuss the physical origins of a spectral peak shift and also a peak broadening.
We find that a spectral broadening of superfluid atoms can be determined from number fluctuations of atoms, while that of normal-state atoms is mainly attributed to quantum fluctuations resulting from hopping of atoms.
To calculate spectra at finite temperatures, based on the sum rule approach, we provide a two-mode approximation assuming that spectra of the superfluid and normal state atoms can be calculated separately.
Our method can properly deal with multi-peak structures of spectra resulting from thermal fluctuations and also coexisting of the superfluid and the normal states.
By combining the two-mode approximation with a finite temperature Gutzwiller approximation, we calculate spectra at finite temperatures by considering realistic systems, and the calculated spectra show nice agreements with those in experiments.
%We discuss differences between the microwave and the laser spectroscopy, and also clarify the lattice-depth and temperature dependences of spectra based on the numerical simulations.
\end{abstract}
\maketitle

\section{Introduction}
%%%===%%%  i1
Ultracold atoms in an optical lattice allows us to simulate quantum phase transitions of lattice fermions and also bosons \cite{Review_Jaksch,Review_Bloch}.
In fact, the superfluid (SF) to the Mott insulator (MI) transition of bosonic atoms has been demonstrated by using various measurement techniques
\cite{SFtoMott_Greiner:Nat,Phase_Gerbier:PRL,NoiseCorr:Nat,in-situ_Mott:Nat,singleSiteHarvard:Sci,singleSiteMPQ:Nat,RF_Gerbier:PRL,Campbell,2009:Bragg:Inguiscio,2010:Sengstock:Bragg,2010:Bragg3D:Heinzen,LatticeMod:PRL,YbMott:PRA}.
The signature of the phase transitions can be observed in certain thermodynamic quantities
\cite{SFtoMott_Greiner:Nat,Phase_Gerbier:PRL,NoiseCorr:Nat,in-situ_Mott:Nat,singleSiteHarvard:Sci,singleSiteMPQ:Nat,YbMott:PRA}.
One of examples is to characterize the transition by observing the disappearance of a coherent peak structure in the number distribution of atoms in the momentum space
\cite{SFtoMott_Greiner:Nat,Phase_Gerbier:PRL,YbMott:PRA}. %-${\bf k}$ space $\langle \hat{n}_{\bf k}\rangle$ \cite{SFtoMott_Greiner:Nat}.
%Other ones are the number distribution in real space \cite{in-situ_Mott:Nat,singleSiteHarvard:Sci,singleSiteMPQ:Nat}, and the density-density correlations \cite{NoiseCorr:Nat}.
A spectroscopic measurement is another useful tool to detect phase transitions
\cite{2009:Bragg:Inguiscio,2010:Sengstock:Bragg,2010:Bragg3D:Heinzen,LatticeMod:PRL,RF_Gerbier:PRL,Campbell}.
This is because much information is included in spectra that reflect the dynamical response of many-body systems after excitations caused by a certain external field.
Furthermore, when the external field is very weak and perturbative, the dynamical response can be connected to thermodynamic quantities of thermal equilibrium states before excitations.
In condensed matter physics, such a relationship, {\it e.g.}, fluctuation-dissipation theorem, has been used to discuss quantum many-body phenomena.
It is thus required to deeply discuss such spectroscopic relationships specific to cold-atom systems.

%%%===%%%  i3
One of pioneering studies on spectroscopic measurement on atoms in a lattice is microwave spectroscopy experiments, where the Mott shell structure has been observed by spectroscopically distinguishing the different number states of atoms \cite{Campbell}.
Theoretically, the corresponding spectra have been studied with an approximation satisfying the (first order) spectral sum rule \cite{Hazzard,Hazzard2010,Yamashita}, which is derived from the general principle of spectroscopy \cite{Oktel}.
The first-order sum rule determines the relationship between the spectral peak position and the two-body correlation function of atoms \cite{Hazzard}. This is a prominent example that connects thermodynamics to dynamics in cold atom systems.
This calculation assumed that the system is at zero temperature, while the realistic experiments have been done at low but finite temperatures.  %, and the first order approximation may be insufficient to investigate the finite temperature spectra.
In addition, such a first order approximation is insufficient to discuss important properties of spectra, such as, a standard deviation and a spectral broadening, which can be connected to fluctuations of atoms in thermal equilibrium.
On the other hand, the laser spectroscopy is now being established \cite{NJP:507,3P2MRI:App,YamashitKato}. %and is applied to the Bose-Hubbard system
The laser and the microwave spectroscopy are understood as a similar type spectroscopy based on electromagnet-field excitations.
However, the laser spectroscopy cannot be straightforwardly described by the formulation of the previous studies \cite{Hazzard}.
This is mainly due to the difference in wavelengths of the external fields.
A reliable theoretical method for comprehensively analyzing these spectroscopy at finite temperatures is now required.

%%%===%%%  i4
In this paper, we theoretically discuss a common formulation for the {microwave} and the laser spectroscopy of ultracold bosonic atoms in a three-dimensional optical lattice.
We start with analyzing the sum rules in the same way as the previous work \cite{Hazzard}, while we extend the approximation to the second order.
This approach allows us to clarify that number fluctuations of atoms in thermal equilibrium can be connected to a broadening of spectra.
%dissipations in dynamics that cause  %, and it further provides
Phenomenological discussions based on the sum rule approach allow us to establish a method for calculating spectra at finite temperatures.
We propose a two-mode approximation assuming that the spectra of condensed SF atoms and uncondensed normal state (NS) atoms are separately dealt with.
The multi-peak structures resulting from thermal fluctuations and also the coexisting of the SF and NS atoms can be appropriately taken account. % thanks to the sum-rule-based formulation.
Using this approximation combined with a finite temperature Gutzwiller approximation \cite{2011:Sugawa:DualMott}, we numerically calculate the microwave and the laser spectra by considering realistic experimental parameters \cite{Campbell,ExpPaper}.
We find that our approximations reproduce essential features of spectra seen in the microwave experiments \cite{Campbell},
and we predict spectra of the realistic laser spectroscopy experiments \cite{ExpPaper}.

\section{Theory of Spectroscopy}\label{sec_theory}
This section is devoted to the general theoretical framework of spectroscopy.
We first explain the model Hamiltonian, and then,
we show the common formulation to describe the microwave and the laser spectroscopy.
To capture essence of the present spectroscopy, we discuss physical properties of spectra in simple model cases.
For simplicity, we set $\hbar=1$ and $k_B=1$. % in this and the next sections.

\subsection{Model Hamiltonian}\label{sec_model}
Before spectroscopic excitations, thermal equilibrium properties of atoms in an optical lattice is well described by the following single-band Bose-Hubbard Hamiltonian \cite{BH:PRB,Jaksch:PRL}:
\begin{eqnarray}
\hat{\cal H}_{g}=-J_{g}\sum_{\langle i,j\rangle}&&\hat{c}_{g,{i}}^\dag\hat{c}_{g,{j}}+\sum_i (V_{g,i}-\mu) \hat{n}_{g,i}\nonumber\\
&&+\frac{U_{g,g}}{2}\sum_i\hat{n}_{g,i}(\hat{n}_{g,i}-1), \label{eq_Hmi}
\end{eqnarray}
where $\hat{c}_{g,{i}}^\dag$ ($\hat{c}_{g,{i}}$) is the creation (annihilation) operator of an unexcited atom at the $i$\,th site, and $\hat{n}_{g,i}$ is the corresponding number operator.
Here, contributions of higher orbitals can be neglected when we consider the low energy properties.
We note that higher orbitals have a role in the spectroscopy as discussed later soon.
%%%===%%%  m2
The Hubbard parameters, i.e., the interaction strength $U_{g,g}$ and the hopping integral $J_{g}$, are evaluated by the {\it ab initio} calculations based on the second quantization using experimental parameters: a lattice constant $a_L$ and a lattice depth $V_0$, which are determined from a wavelength and an intensity of the lattice laser, respectively, and $a_{g,g}$ a scattering length between two unexcited atoms. %${\rm {}^1S_0}$ atoms.
The chemical potential $\mu$ is determined so as to fix the total number of atoms $N_{\rm tot}=\sum_i \langle \hat{n}_{g,i}\rangle$ and  $V_{g, i}$ is the trapping potential.
In the following, for simplicity, we omit to explicitly write down $\mu$, which can be included in a global shift of $V_{g,i}$.
%, where $\langle \cdots \rangle$ denotes the statistical average at thermal equilibrium,

\subsection{Spectroscopy}\label{sec_spec}
%%%===%%%  m1
In the microwave and the laser spectroscopy, excitation processes caused by an external electro-magnetic field are generally described by $\Gamma \hat{\cal O}_{\rm ex} e^{i\omega t+i {\bf K}_{\rm ex}\cdot \hat{\bf r}}$ \cite{Hazzard}, where $\Gamma$,  $\omega$, and ${\bf K}_{\rm ex}$ are a non-dimensional normalized amplitude, an angular frequency, and a wavevector of the external field, respectively, and $t$ and ${\bf r}$ are time and position.
Here $\hat O_{\rm ex}$ is an excitation operator defined as follows.
For convenience, we define $\sum_\alpha \rho_\alpha \hat{\cal O}_{\rm ex,\alpha}$ as a second quantization of $\hat{\cal O}_{\rm ex} e^{i {\bf K}_{\rm ex}\cdot \hat{\bf r}}$, and
%$
%O_{ex}=\sum_{\bf k}p_{{\bf k+k}_{ex}}^\dag s_{\bf k} + {\rm H.c}
%$
\begin{equation}
\hat{\cal O}_{\rm ex,\alpha} = \sum_{i} e^{i{\bf k}_{\rm ex}\cdot {\bf r}_i} \hat{c}_{e\alpha,i}^\dag\hat{c}_{g,{i}}+{\rm H.c.},
\label{eq_oex}
\end{equation}
where $\hat{c}_{e\alpha,i}$ is the annihilation operator of an excited atom in the $\alpha$-th orbital at the $i$ th site of the position ${\bf r}_i$.
Note that ${\bf k}_{\rm ex}$ is a reduced wavevector in the first Brillouin zone defined by ${\bf k}_{\rm ex}\equiv {\bf K}_{\rm ex} + {\bf G}$, where ${\bf G}$ represents any reciprocal vectors with $e^{i{\bf G}\cdot{\bf r}_i}=1$.
An excitation matrix $\rho_\alpha$ is defined by
\begin{equation}
\rho_\alpha =\int d{\bf r}W^*_\alpha({\bf r-r}_i) e^{i{\bf K}_{\rm ex}\cdot ({\bf r-r}_i)} W_1({\bf r-r}_i),
\label{eq_rhoalpha}
\end{equation}
where $W_\alpha({\bf r}-{\bf r}_i)$ is the $\alpha$-th Wannier orbital at the $i$th site for the excited atoms, and $W_1({\bf r-r}_i)$ is that for the unexcited atoms (i.e., $\alpha=1$).
Here, $|\rho_\alpha|^2$ represents the probability that the orbital of atoms changes from the lowest to the $\alpha$-th orbital during excitations.
The orthogonality of the Wannier orbitals assures a condition $\sum_\alpha |\rho_\alpha|^2=1$.
In this paper, we neglect the probability that atoms are excited to the different lattice sites (i.e., inter-site excitation), because it is exponentially smaller than that of the onsite excitations.
Namely, we assume that $\rho_\alpha^{i,j}=\int d{\bf r}W^*_\alpha({\bf r-r}_j) e^{i{\bf K}_{\rm ex}\cdot ({\bf r-r}_i)} W_1({\bf r-r}_i)$ vanishes except for $i=j$.

We focus on the weak excitation limit under the condition of $|\Gamma|\ll1$.
The excitation spectra can be formally given by $I(\omega)= \sum_\alpha |\rho_\alpha|^2 I_\alpha(\omega)$, and
\begin{equation}
I_\alpha(\omega)=|\Gamma|^2 \sum_{n',n} |\langle n'|{\hat O}_{{\rm ex},\alpha}|n\rangle|^2
{e}^{-(E_{n}-\Omega)/T} \delta(\omega-E_n'+E_n), \label{eq_gen_spc}
\end{equation}
where $|n \rangle$ is the eigenstate of Hamiltonian ${\cal H}_g$ in Eq.\,(\ref{eq_Hmi}) with energy $E_n$, and $\Omega(=-T \ln \sum_n e^{-E_{n}/T})$ is the grand potential.
Note that the conservation law of the number of excited atoms allows us to decompose $I(\omega)$ into the sum of $I_\alpha(\omega)$.
The excited state $|n'\rangle$ is the eigenstate of Hamiltonian $\hat{\cal H}\equiv \hat{\cal H}_{g}+\hat{\cal H}_e+\hat {\cal H}_{ge}$, and $E_n'$ is its energy.
Here $\hat{\cal H}_e$ and $\hat{\cal H}_{ge}$ are given by
\begin{eqnarray}
\hat {\cal H}_{e}&=& -\!\!\!\!\sum_{\langle i,j \rangle,\alpha}J_{e\alpha} \hat{c}^\dag_{e\alpha,i} \hat{c}_{e\alpha,j}
+ \sum_{i,\alpha} (\Delta_{e\alpha}+V_{e\alpha,i}) \hat{n}_{e\alpha,i}, \\
\!\!\hat {\cal H}_{ge}&=& \sum_{i,\alpha} U_{g,e\alpha} \hat{n}_{e\alpha,i}\hat{n}_{g,i},
\end{eqnarray}
where $J_{e\alpha}$ is the hopping integral of excited atoms in the $\alpha$-th orbital, and $U_{g,e\alpha}$ is the onsite interaction between the first orbital unexcited and the $\alpha$ th orbital excited atoms.
Note that the interaction between two excited atoms $U_{e\alpha,e\beta}$ can be reasonably neglected in the limit of weak excitations.
$\Delta_{e\alpha}$ represents the energy difference between the unexcited atoms in the lowest orbital and the excited atoms in the $\alpha$ th orbital.
We can always set $\Delta_{e1}=0$ by appropriately choosing the origin of the spectral frequency.
The spectral intensity proportional to $|\Gamma|^2$ is determined so as to satisfy the integral condition $\int I(\omega) d\omega= const.$, when we compare our analyses with the experimental observations.
Thus, we can neglect the quantitative aspect of $\Gamma$ by setting $\Gamma=1$ without loss of generality.

\subsection{Sum rules}\label{sec_sumrule}
%%%===%%%  t1
We discuss the moment expansions of the spectral function by following the previous studies \cite{Hazzard,Hazzard2010,Yamashita}. %and then we point out that several terms neglected in the previous studies play an important role in laser spectroscopy.
In general, spectra given by Eq.\,(\ref{eq_gen_spc}) should satisfy the following sum rules in terms of  $M_\alpha^{(n)}$ the $n$-th order moment:
\begin{eqnarray}
\nonumber M^{(n)}_\alpha&\equiv&\int d\omega \, \omega^n I_\alpha(\omega),\\
&=& \langle [[[\hat{O}_{\rm ex,\alpha}^\dag, \hat{\cal H}],\hat{\cal H},\cdots] \hat{O}_{\rm ex,\alpha}\rangle, \label{eq_gen_sum}
\end{eqnarray} %\sum_\alpha |\rho_\alpha|^2
where $[[[\hat{O}_{\rm ex,\alpha}^\dag, \hat{\cal H}],\hat{\cal H},\cdots]$ denotes the $n$-times commutator between $\hat{O}_{\rm ex,\alpha}$ and $\hat{\cal H}$.
These relations indicate that certain statistic quantities in thermal equilibrium have a relation with some properties of spectral functions reflecting a dynamical response of the system.
For examples, by considering up to the second order moments, the spectral mean value $\bar{\omega}_\alpha$ is given by $\bar{\omega}_\alpha=M_\alpha^{(1)}/M_\alpha^{(0)}$, %%\int I(\omega) \omega d\omega/\int I(\omega) d\omega=
and the standard deviation $\sigma_\alpha$ can be written as $\sigma^2_\alpha=M_\alpha^{(2)}/M_\alpha^{(0)}-(M_\alpha^{(1)}/M_\alpha^{(0)})^2$.
Naively, we can stress that $\bar{\omega}_\alpha$ and $\sigma_\alpha$ determine a spectral peak position and its width, respectively, and the relations defined by Eq.\,(\ref{eq_gen_sum}) can describe the physical origin of the peak shift and also broadening caused by the many-body effects.
In what follows,  to discuss these important spectral properties, we analyze the sum rules up to the second order.

We can derive the following expressions with respect to the corresponding sum rules.
The zeroth order is given by
\begin{eqnarray}
M^{(0)}_\alpha= \langle \hat{O}_{\rm ex,\alpha}^\dag \hat{O}_{\rm ex,\alpha} \rangle = \sum_i \langle \hat{n}_{g,i}\rangle= N_{\rm tot}.
\label{eq_mom_zth}
\end{eqnarray}
The first order moment is written as
\begin{eqnarray}
M^{(1)}_\alpha&=&\langle [\hat{O}_{\rm ex,\alpha}^\dag, {\cal H}] \hat{O}_{\rm ex,\alpha} \rangle \nonumber\\
&=& (U_{g,e\alpha}-U_{g,g}) \sum_i \langle \hat{c}_{g,i}^\dag \hat{c}_{g,i}^\dag \hat{c}_{g,i} \hat{c}_{g,i} \rangle \nonumber\\
&+& \sum_i(V_{e\alpha,i}-V_{g,i}+\Delta_{e\alpha})\langle \hat{n}_{g,i}\rangle \nonumber\\
&+& \sum_{i}\sum_{\bf d}(J_g-J_{e\alpha} e^{i{\bf k}_{\rm ex}\cdot{\bf d}})  \langle \hat{c}_{g,i}^\dag \hat{c}_{g,i+d} \rangle, \label{eq_mom_fst}
\end{eqnarray}
where $\sum_{\bf d}$ represents a summation over the adjacent sites, and the second order is
\begin{widetext}
\begin{eqnarray}
M^{(2)}_\alpha&=&\langle[ [\hat{O}_{\rm ex,\alpha}^\dag, {\cal H}], {\cal H}] \hat{O}_{\rm ex,\alpha} \rangle \nonumber\\
&=&  (U_{g,e\alpha}-U_{g,g})^2 \sum_i \langle \hat{c}_{g,i}^\dag \hat{c}_{g,i}^\dag \hat{c}_{g,i}^\dag \hat{c}_{g,i} \hat{c}_{g,i} \hat{c}_{g,i} \rangle
+ (U_{g,e\alpha}-U_{g,g})^2 \sum_i \langle \hat{c}_{g,i}^\dag  \hat{c}_{g,i}^\dag \hat{c}_{g,i} \hat{c}_{g,i} \rangle \nonumber\\
&&+\sum_i (V_{e\alpha,i}-V_{g,i}+\Delta_{e\alpha})^2 \langle \hat{n}_i\rangle
+ \sum_i 2(U_{g,e\alpha}-U_{g,g})(V_{e\alpha,i}-V_{g,i}+\Delta_{e\alpha}) \langle \hat{c}_{g,i}^\dag  \hat{c}_{g,i}^\dag \hat{c}_{g,i} \hat{c}_{g,i} \rangle \nonumber\\
&&+\sum_{i} \sum_{\bf d}2(J_g-J_{e\alpha} e^{i{\bf k}_{\rm ex}\cdot{\bf d}})(U_{g,e\alpha}-U_{g,g})  \langle \hat{c}_{g,i}^\dag \hat{c}_{g,i} \hat{c}_{g,i}^\dag\hat{c}_{g,i+d} \rangle \nonumber \\
&&+\sum_{i} \sum_{\bf d}2(J_g-J_{e\alpha} e^{i{\bf k}_{\rm ex}\cdot{\bf d}})(V_{e\alpha,i}-V_{g,i}+\Delta_{e\alpha})  \langle \hat{c}_{g,i}^\dag \hat{c}_{g,i+d} \rangle \nonumber\\
&&+\sum_{i} \sum_{\bf d,d'} (J_g-J_{e\alpha} e^{i{\bf k}_{\rm ex}\cdot {\bf d}})(J_g-J_{e\alpha} e^{i{\bf k}_{\rm ex}\cdot {\bf d}'})
\times(1-\delta_{{\bf d+d}',{\bf 0}})\langle \hat{c}_{g,i}^\dag \hat{c}_{g,i+d+d'}\rangle \nonumber\\
&&+\sum_{i}  \sum_{\bf d}|J_g-J_{e\alpha} e^{i{\bf k}_{\rm ex}\cdot{\bf d}}|^2  \langle \hat{n}_{g,i} \rangle.
\label{eq_mom_snd}
\end{eqnarray}
\end{widetext}

The expressions in Eqs.\,(\ref{eq_mom_zth})-(\ref{eq_mom_snd}) include the onsite multi-body correlation functions $G_{\ell,i} \equiv \langle (\hat{c}_{g,i}^\dag)^\ell (\hat{c}_{g,i})^\ell \rangle$, where $G_{1,i}$ is equivalent to the averaged number of atoms $n_i\equiv \langle\hat{n}_{g,i}  \rangle$.
% defined as follows:  one-body $G_{1,i}\equiv\langle\hat{n}_{g,i}  \rangle$, which is equivalent to the average number of atoms $n_i$, two-body $G_{2,i}\equiv \langle \hat{c}_{g,i}^\dag \hat{c}_{g,i}^\dag  \hat{c}_{g,i} \hat{c}_{g,i} \rangle$, and three-body correlations $G_{3,i}\equiv\langle \hat{c}_{g,i}^\dag \hat{c}_{g,i}^\dag \hat{c}_{g,i}^\dag \hat{c}_{g,i} \hat{c}_{g,i} \hat{c}_{g,i} \rangle$.
The hopping Hamiltonian yields the inter-site correlations such as ${\cal G}_{i,j}\equiv \langle \hat{c}_{g,i}^\dag \hat{c}_{g,j} \rangle$, where ${\cal G}_{i,i}=n_i$, and ${\cal G}_{2,i,j}\equiv \langle \hat{c}_{g,i}^\dag \hat{c}_{g,i} \hat{c}_{g,i}^\dag\hat{c}_{g,j} \rangle$.
In the second (or higher) order of moment, the hopping Hamiltonian yields also the onsite one-body correlation $G_{1,i}(=n_i)$ [see the last term in Eq.\,(\ref{eq_mom_snd})].
Note that this type of terms is caused by a round-trip hopping process.
These terms are proportional to $J_g-J_{e\alpha} e^{i{\bf k}_{\rm ex}\cdot {\bf d}}$, where $e^{i{\bf k}_{\rm ex}\cdot {\bf d}}$ describes the momentum transfer from the external field to atoms.
The momentum transfer can be regarded as a back action of the measurements, which plays an important role in spectra.

\subsection{Physical meaning of the spectral deviations and spectral mean value}
\label{sec_discuss}

It is convenient to discuss the physical meaning of spectra, which can be figured out from the sum-rule approach.
The spectral mean value $\bar{\omega}_\alpha$ can be written as the sum of $\delta{U}_\alpha$, $\delta{V}_\alpha$, $\delta{\Delta}_\alpha$ and $\delta{J_\alpha}$, which are spectral energy shifts caused by the effects of interaction, trapping potential, band gap, and hopping, respectively.
As discussed in the previous study \cite{Hazzard}, $\delta{U}_\alpha \equiv (U_{g,e\alpha}-U_{g,g}) \sum_i G_{2,i} /N_{\rm tot}$ is a collisional energy shift. %, which corresponds to the mean-field energy shift.
Two terms, $\delta{V}_\alpha\equiv \sum_i (V_{e\alpha,i}-V_{g,i}) {n}_{i}/N_{\rm tot} $ and $\delta{\Delta}_\alpha \equiv \sum_i \Delta_{e\alpha}{n}_{i}/N_{\rm tot}$, are related to the statistical average of the number of atoms.
Here, $\delta{\Delta}_\alpha$ reduces to a constant $\Delta_{e\alpha}$ having no connection to any thermodynamic quantities, while $\delta{V}_\alpha$ describes the effects of the inhomogeneity of the system.
%The hopping energy shift $\delta{J}_\alpha \equiv \sum_{i}\sum_{\bf d}(J_g-J_{e\alpha} e^{i{\bf k}_{\rm ex}\cdot {\bf d}})  \langle \hat{c}_{g,i}^\dag \hat{c}_{g,i+d} \rangle/N_{\rm tot}$ is related to the inter-site correlation.
The hopping energy shift $\delta{J}_\alpha \equiv \sum_{i}\sum_{\bf d}(J_g-J_{e\alpha} e^{i{\bf k}_{\rm ex}\cdot {\bf d}})  {\cal G}_{i,i+d}/N_{\rm tot}$ is related to the inter-site correlation.
The Bloch band picture makes physical meanings of this term clear.
We define the kinetic energy shift  $\delta{K}_\alpha\equiv \delta{J}_\alpha+\delta{\Delta}_\alpha$, which can be rewritten as $\sum_{\bf k} (\varepsilon_{e\alpha,{\bf k+k}_{\rm ex}}-\varepsilon_{g,{\bf k}}) n_{g,{\bf k}}/N_{\rm tot}$, where $n_{g,{\bf k}}=\sum_{i,j}e^{i{\bf k}\cdot({\bf r}_i-{\bf r}_j)} {\cal G}_{i,j}$ is the momentum space distribution, and $\varepsilon_{g,\bf k}$ and $\varepsilon_{e \alpha,\bf k}$ are the dispersions of the unexcited atoms in the lowest orbital and of the excited atoms in the $\alpha$-orbital, respectively.
Because of the momentum conservation law, a wavevector ${\bf k}_{\rm ex}$ is transfered from an external field to atoms.
It turns out that $\delta{K}_\alpha$ becomes large for a large ${\bf k}_{\rm ex}$.
Note that the discrete translational symmetry imposes ${\bf k}_{\rm ex}$ to be a reduced wavevector in the first Brillouin zone.
%\langle \hat{c}^\dag_{g,i}\hat{c}_{g,j} \rangle

We next discuss the standard deviation $\sigma_\alpha$ related to the second order moment as $\sigma^2_\alpha=M_\alpha^{(2)}/M_\alpha^{(0)}-(M_\alpha^{(1)}/M_\alpha^{(0)})^2$.
Roughly speaking, the standard deviation characterizes broadening of spectra.
For clear vision, we now focus on a uniform system by setting $V_{e\alpha,i}-V_{g,i}=0$.
Considering the physical origin of the deviation, we can rewrite $\sigma_\alpha$ as $\sigma^2_\alpha\equiv\sigma^2_{U, \alpha}+\sigma^2_{K, \alpha}+\sigma^2_{\sqrt{UK}, \alpha}$, where
the deviation induced by the correlations $\sigma_{U, \alpha}$ is defined as $\sigma_{U, \alpha}^2\equiv (U_{g,e\alpha}-U_{g,g})^2[\sum_i(G_{3,i}+G_{2,i})/N_{\rm tot}-(\sum_i G_{2,i}/N_{\rm tot})^2]$, and that caused by kinetic terms $\sigma_{K, \alpha}$ is defined as $\sigma_{K, \alpha}^2\equiv \sum_{\bf k}(\varepsilon_{g,{\bf k}}-\varepsilon_{e\alpha,{\bf k+k}_{\rm ex}})^2 n_{g,{\bf k}}/N_{\rm tot} -[\sum_{\bf k} (\varepsilon_{e\alpha,{\bf k+k}_{\rm ex}}-\varepsilon_{g,{\bf k}}) n_{g,{\bf k}}/N_{\rm tot}]^2$, and that originating from the cross terms $\sigma_{\sqrt{UK}, \alpha}$ is given by $\sigma_{\sqrt{UK}, \alpha}^2\equiv 2(U_{g,e\alpha}-U_{g,g})\sum_{\bf d}(J_{g}-J_{e\alpha} e^{i{\bf k}_{\rm ex}\cdot {\bf d}})[\sum_i {\cal G}_{2,i,i+d}/N_{\rm tot} - (\sum_i G_{2,i}/N_{\rm tot})(\sum_i {\cal G}_{i,i+d}/N_{\rm tot})]$.
%the deviation induced by the correlations $\sigma_{U, \alpha}$ is defined as $\sigma_{U, \alpha}^2\equiv (U_{g,e\alpha}-U_{g,g})^2[\sum_i(G_{3,i}+G_{2,i})/N_{\rm tot}-(\sum_i G_{2,i}/N_{\rm tot})^2]$, and that caused by kinetic terms $\sigma_{K, \alpha}$ is defined as $\sigma_{K, \alpha}^2\equiv \sum_{\bf k}(\varepsilon_{g,{\bf k}}-\varepsilon_{e\alpha,{\bf k+k}_{\rm ex}})^2 n_{g,{\bf k}}/N_{\rm tot} -[\sum_{\bf k} (\varepsilon_{e\alpha,{\bf k+k}_{\rm ex}}-\varepsilon_{g,{\bf k}}) n_{g,{\bf k}}/N_{\rm tot}]^2$, and that originating from the cross terms $\sigma_{\sqrt{UK}, \alpha}$ is given by $\sigma_{\sqrt{UK}, \alpha}^2\equiv 2(U_{g,e\alpha}-U_{g,g})\sum_{\bf d}(J_{g}-J_{e\alpha} e^{i{\bf k}_{\rm ex}\cdot {\bf d}})[\sum_i \langle \hat{c}_{g,i}^\dag \hat{c}_{g,i} \hat{c}_{g,i}^\dag\hat{c}_{g,i+d} \rangle/N_{\rm tot} - (\sum_i G_{2,i}/N_{\rm tot})(\sum_i \langle \hat{c}_{g,i}^\dag\hat{c}_{g,i+d} \rangle/N_{\rm tot})]$.

%To clarify physical meanings of the complex form of $\sigma_\alpha$, we further need to simplify the problem as follows.
In the following, we consider two simplified model cases and calculate a mean value $\bar{\omega}_\alpha$ and a deviation $\sigma_\alpha$, so as to discuss what are the physical origins of spectral peak shifts and broadening.
%the physical meanings of
%$\sigma_\alpha$ and $\bar{\omega}_\alpha$.
Here we focus on uniform systems at zero temperature for simplicity, and inhomogeneous systems at finite temperatures will be discussed in Sec.\ \ref{sec_method}.
We first consider SF states at zero temperature and use the following simple mean-field approximation.
We assume that $\langle \hat{c}_{g,i} \rangle$ is finite and is given by a classical complex number $c_{i}$, which leads to $G_{\ell,i} \sim |c_{i}|^{2\ell}$, $\langle \hat{c}_{g,i}^\dag \hat{c}_{g,j} \rangle \sim c^*_{i}c_{j}$, and
 $n_{g,{\bf k}} \sim \delta_{{\bf k},{\bf k}_{0}} N_{\rm SF}$, where $N_{\rm SF}$ is the number of the condensed SF atoms, and ${\bf k}_{0}$ is a wavevector at the bottom of a dispersion $\varepsilon_{g,{\bf k}}$ (usually ${\bf k}_0={\bf 0}$ for a positive $J_{g}$).
We here also assume that almost all atoms are in the condensed state $N_{\rm SF} \sim N_{\rm tot}$.
The spectral mean value is now given by $\bar{\omega}_\alpha = (U_{g,e\alpha}-U_{g,g}) {n}_i + (\varepsilon_{e\alpha,{\bf k}_0+{\bf k}_{\rm ex}}-\varepsilon_{g,{\bf k}_0})$. %, where $\bar{n} (\equiv \sum_i n_i/N_{\rm tot})$ is the averaged number of atoms in each site.
It is consistent with the previous study \cite{Hazzard}, while the additional kinetic energy shift is found.
The deviation now reduces to $ \sigma^2_\alpha = (U_{g,e\alpha}-U_{g,g})^2 {n}_i (=\sigma_{U,\alpha}^2)$.
Interestingly, we find that the deviation has a contribution of an interaction term $\sigma_{U,\alpha}$ only.
Here because of the subtraction in $M_\alpha^{(2)}/M_\alpha^{(0)}-(M_\alpha^{(1)}/M_\alpha^{(0)})^2$, the deviations $\sigma_{K, \alpha}$ and $\sigma_{\sqrt{KU},\alpha}$ are canceled out under the pure-condensation condition $n_{g,{\bf k}} \sim \delta_{{\bf k},{\bf k}_{0}} N_{\rm tot}$.
The number of the coherent SF atoms is indefinite, and number fluctuations $\Delta n_i$ of such a coherent state are written as $\sqrt{{n}_i}$, where $(\Delta n_i)^2=\langle \hat{n}^2_{g,i} \rangle- {n}_i^2$.
Thus, we can conclude that the spectral deviation of coherent SF states is connected to number fluctuations in thermal equilibrium as $\sigma_\alpha=|U_{g,e\alpha}-U_{g,g}| \Delta n_i$.

We next consider MI states with $m$ atoms in each site at zero temperature.
For MI states, it is reasonable to set inter-site correlations ${\cal G}_{i,j}$ for $i\not=j$ to be zero.
The spectral mean value is now given by $\bar{\omega}_\alpha = (U_{g,e\alpha}-U_{g,g})(m-1)+\Delta_{e\alpha}$, which is equivalent to the previous study \cite{Hazzard}.
%This indicates that the present spectroscopy allows us to distinguish the different number states of atoms when $|U_{g,e\alpha}-U_{g,g}|$ is large enough \cite{Campbell,Hazzard}.
The spectral deviation reduces to $\sigma_\alpha^2= {\sum_{\bf d} |J_g-J_{e\alpha}e^{i{\bf k}_{\rm ex}\cdot {\bf d}}|^2} (=\sigma_{K,\alpha}^2)$.
In the same manner as the above, the correlation-induced deviation $\sigma_{U,\alpha}$ cancels out as $\sigma_{U,\alpha}^2=m(m-1)(m-2)/m+m(m-1)/m-[m(m-1)/m]^2=0$.
The cross-term-induced deviation $\sigma_{\sqrt{UK},\alpha}$ is also zero because of negligible inter-site correlations.
For MI atoms at zero temperature, the number of atoms is definite, and there are no number fluctuations $\Delta n_i=0$.
On the other hand, the phase and the momentum is indefinite, and thus a spectral broadening of MI atoms is caused by kinetic fluctuations described by $\sigma_{K,\alpha}$.
By executing the ${\bf k}$-summation in $\sigma_{K,\alpha}$ with a constant momentum distribution $n_{g, {\bf k}}(=m)$ of uniform MI states, we obtain $\sigma_{K,\alpha}^2={\sum_{\bf d} |J_g-J_{e\alpha}e^{i{\bf k}_{\rm ex}\cdot {\bf d}}|^2}$.
This deviation can be connected to quantum fluctuations resulting from the round-trip hopping process given in the last term in Eq.\,(\ref{eq_mom_snd}).

As demonstrated by the previous experiments \cite{Campbell}, the present spectroscopy has an ability to distinguish the different number states, when $|U_{g,e\alpha}-U_{g,g}|$ is large enough.
This feature of spectra can be explained by the first order sum rule, $\bar{\omega}_\alpha = (U_{g,e\alpha}-U_{g,g})(m-1)$, as discussed in the previous study \cite{Hazzard}.
The above second-order sum rule approach further indicates that number fluctuations of atoms in thermal equilibrium $\Delta n_i$ play an important role in such number-resolving spectroscopy.
%In fact, spectral mean values $\bar{\omega}_\alpha$ of SF (MI) atoms at zero temperature are characterized by the averaged (definite) number of atoms \cite{Hazzard}.
In fact, a spectral deviation $\sigma_\alpha$ of coherent SF atoms can be determined from number fluctuations: $\sigma_\alpha=|U_{g,e\alpha}-U_{g,g}| \Delta n_i$.
Even for MI atoms with $\Delta n_i=0$, a spectral deviation is finite owing to kinetic fluctuations, which is attributed to the indefinite phase and momentum in a reflection of the definite number and position of MI atoms.
% $\sigma_{K,\alpha}=\sqrt{\sum_{\bf d} |J_g-J_{e\alpha}e^{i{\bf k}_{\rm ex}\cdot {\bf d}}|^2}$
This deviation of the number definite states is not related to any thermodynamic quantities, and thus this constant $\sigma_\alpha$ of $\sqrt{\sum_{\bf d} |J_g-J_{e\alpha}e^{i{\bf k}_{\rm ex}\cdot {\bf d}}|^2}$ is the intrinsic lower limit of a spectral linewidth (see Sec.\ \ref{sec_binary}).
On the other hand, the deviations of general states with both phase and number fluctuations are given by the summation, $\sigma^2_{U, \alpha}+\sigma^2_{K, \alpha}+\sigma^2_{\sqrt{UK}, \alpha}$.
We find that the spectral deviations of a specific ground state (SF and MI) with a definite quantity (phase and number) are characterized by fluctuations resulting from the conjugate indefinite quantity (number and phase, respectively).
It should be noted that the second-order sum rule makes clear the fact that the spectral measurements are governed by the uncertainty principle.

%We should note that
These physical properties of spectra mentioned above are figured out from the general principle of weak excitation spectroscopy with given Hamiltonian $\hat{\cal H}$ and an operator $\hat{\cal O}_{{\rm ex},\alpha}$.
For example, an excitation operator $\hat{\cal O}_{{\rm ex},\alpha}$ characterizes the intrinsic spectral broadening of $\sqrt{\sum_{\bf d} |J_g-J_{e\alpha}e^{i{\bf k}_{\rm ex}\cdot {\bf d}}|^2}$, where a momentum transfer of ${\bf k}_{\rm ex}$, which is a back action of the measurements, determines a quantitative aspect of a spectral width.
Bose-Hubbard Hamiltonian ${\cal H}_g$ includes kinetic and interaction terms, and the competition between these two conjugate terms is the origin of the SF-MI transitions.
The second-order sum rule approach clarifies that spectral deviations reflect the completely different properties of these two conjugate states, SF and MI.
We can thus conclude that the present spectroscopy will be a sensitive tool for detecting the SF-MI transitions.
Note that the first order sum rule approach is insufficient to clarify these important features of spectroscopy.
%The second-order sum rule approach clarifies that
%allows us to connect fluctuations in thermal equilibrium to dissipations of the dynamical response of atoms that cause spectral broadening.
%quantum phase transitions.
%completely different properties of spectra between SF and MI states allow us to characterize the SF-MI transitions, suggesting that
However,
the above simplified discussions cannot be straightforward applied to the finite temperature spectra. % of the inhomogeneous systems.
In the next section, we thus propose a two-mode approximation to numerically calculate spectra that satisfy the sum rules.

\section{Methods}\label{sec_method}

In this section, we provide a numerical method for calculating spectra at finite temperatures.
We first explain the finite temperature Gutzwiller approximation \cite{2011:Sugawa:DualMott}, which allows us to efficiently obtain the thermodynamic quantities in Eqs.\,(\ref{eq_mom_zth})-(\ref{eq_mom_snd}).
We next provide a two-mode approximation  to numerically calculate finite temperature spectra in inhomogeneous systems.
At the end of this section, we compare our method with the previous formulations \cite{Hazzard,Hazzard2010}.

\subsection{Finite temperature Gutzwiller approximation}\label{sec_gutz}
The Gutzwiller approximation allows us to efficiently analyze the thermal equilibrium properties described by the Bose-Hubbard Hamiltonian in Eq.\,(\ref{eq_Hmi}). % \cite{2011:Sugawa:DualMott}
This is a mean-field approximation considering up to the first-order collection in terms of $J_{g}$ and well describes the SF-MI transitions in high dimensional systems.
Here, ${\cal \hat{H}}_{g}$ is then approximated by a set of the effective local Hamiltonian ${\cal \hat{H}}_{\rm loc}=\sum_i{\cal \hat{H}}_{{\rm loc},i}$ with
\begin{equation}
{\cal \hat{H}}_{{\rm loc},i}=J^{\rm eff}_{g,i}\hat{c}^\dag_{g,i}+J^{\rm eff *}_{g,i}\hat{c}_{g,i}+V_{g,i}\hat{n}_{g,i}+\frac{U_{g,g}}{2}\hat{n}_{g,i}(\hat{n}_{g,i}-1),
\label{eq_HmiLoc}
\end{equation}
where $J^{\rm eff}_{g,i}$ is determined from a self-consistent condition $J^{\rm eff}_{g,i}=-\sum_{\bf d} J_{g}\langle \hat{c}_{g,i+d} \rangle$.
Using exact diagonalization, we can numerically calculate statistical quantities such as $c_i\equiv \langle \hat{c}_{g,i} \rangle$ at finite temperatures \cite{2011:Sugawa:DualMott}.

As discussed in Sec.\ \ref{sec_discuss}, a finite $c_i$ effectively describes the Bose-Einstein condensates (BEC) within the mean-field approximation.
Here, the number of condensed SF atoms in each site can be defined by $n_{{\rm SF},i}= |c_i|^2$.
Both thermal fluctuations and interactions cause coexisting of condensed SF and uncondensed NS such as MI and normal fluid (NF).
The annihilation operator of NS atoms at the $i$th site is effectively given by  $\hat{c}_{{\rm NS},i}=\hat{c}_{g,i}-c_i$, where NS atoms satisfy always a condition $\langle \hat{c}_{{\rm NS},i}\rangle =0$.
The number of NS atoms $n_{{\rm NS},i}\equiv \langle \hat{c}^\dag_{{\rm NS},i} \hat{c}_{{\rm NS},i} \rangle$ can be written as $n_{{\rm NS},i}={n}_{i} - n_{{\rm SF},i}$.
This leads to the following reasonable condition: $N_{\rm tot}=N_{\rm SF}+N_{\rm NS}$, where $N_{\rm SF (NS)}=\sum_i n_{{\rm SF (NS)},i}$.
The total number of atoms is given by the sum of the total number of SF and NS atoms.

It is useful to briefly explain how to calculate the thermal quantities in the moments Eqs.\,(\ref{eq_mom_fst}) and (\ref{eq_mom_snd}).
The onsite correlation functions $G_{n,i}$ can be calculated straightforwardly by diagonalizing the effective local Hamiltonian ${\cal \hat{H}}_{{\rm loc},i}$.
On the basis of this local approximation, the inter-site correlation is described by ${\cal G}_{i,j}\sim \langle \hat{c}^\dag_{g,i}\rangle \langle \hat{c}_{g,j} \rangle (=c_i^* c_j)$ for $i\not =j$.
The higher-order inter-site correlations ${\cal G}_{2,i,i+d}$ also reduce to $\langle \hat{c}^\dag_{g,i}\hat{c}_{g,i}\hat{c}^\dag_{g,i}\rangle\langle \hat{c}_{g,j} \rangle$.
These expressions mean that
the inter-site correlations of NS atoms $\langle \hat{c}_{{\rm NS},i}^\dag \hat{c}_{{\rm NS},j}\rangle$ for $i\not =j$ are approximately set to be zero, and thus, two kinds of NS states, MI and NF, are dealt with approximately in the same way.
We should note that the MI states appearing at lower temperatures can be characterized by focusing on the creation of the Mott shell structures and also the suppressed entropy per site \cite{2011:Sugawa:DualMott}.
We can thus effectively calculate that thermal fluctuations cause the MI-NF crossover within this local approximation.

\subsection{Two-mode approximation}\label{sec_binary}
%%%===%%%  t1

Next, we provide a two-mode approximation that helps us to calculate spectra at finite temperatures.
We assume that $I_\alpha(\omega)$ in Eq.\,(\ref{eq_gen_spc}) can be decomposed into two components resulting from the contributions of SF and NS atoms:
\begin{equation}
I_\alpha(\omega)=I_\alpha^{\rm SF}(\omega)+I_\alpha^{\rm NS}(\omega).
\end{equation}
Two types of uncondensed NS (MI and NF)  states appear at finite temperatures.
As mentioned in Sec.\ \ref{sec_gutz}, within the Gutzwiller approximation, NF states are approximately dealt with in the same way as MI states based on the local Hamiltonian picture.
As discussed in Sec.\ \ref{sec_discuss}, at zero temperature, spectra of coherent SF atoms show completely different properties by comparing with those of MI atoms.
Note that the special characteristics of spectra of SF atoms result from the phase coherence caused by BEC.
We thus deal with spectra of SF atoms in different way to two types of NS atoms.

On the basis of the two-mode approximation, we reconsider the sum rules for the spectral moments:
\begin{equation}
M^{(n)}_\alpha=\int d\omega\,\omega^n I_{\alpha}^{\rm NS}(\omega)+\int d\omega\,\omega^nI_\alpha^{\rm SF}(\omega).
\nonumber
\end{equation}
The sum rules up to the second order (i.e., up to $n=2$) provide the following relations:
\begin{eqnarray}
M^{(0)}_\alpha&=&N_{\rm SF}+N_{\rm NS}, \label{eq_BFsum_fst}\\
M^{(1)}_\alpha&=&N_{\rm NS}\bar{\omega}_\alpha^{\rm NS}+N_{\rm SF}\bar{\omega}_\alpha^{\rm SF}, \label{eq_BFsum_snd}\\
M^{(2)}_\alpha&=& [({\sigma}_\alpha^{\rm NS})^2+(\bar{\omega}_\alpha^{\rm NS})^2] N_{\rm NS}+
[({\sigma}_\alpha^{\rm SF})^2+(\bar{\omega}_\alpha^{\rm SF})^2] N_{\rm SF} \nonumber, \\\label{eq_BFsum_trd}
%&& \!\!\!\!+(\bar{\omega}_\alpha^{\rm SF}-\bar{\omega}_\alpha^{\rm NS})^2N_{\rm SF}N_{\rm NS}/N_{\rm tot},
\end{eqnarray}
where $\bar{\omega}_\alpha^{\rm NS}$ and $\bar{\omega}_\alpha^{\rm SF}$ are the spectral mean value, and % for the NS and SF spectra, respectively. where
${\sigma}_{\rm NS}$ and ${\sigma}_{\rm SF}$ are the spectral standard deviation for the NS and SF spectra, respectively.
The zeroth order sum rule in Eq.\,(\ref{eq_BFsum_fst}) simply offers the condition associated with the total number of atoms, which is always satisfied within the Gutzwiller treatment as mentioned in Sec.\ \ref{sec_gutz}.
On the other hand, the first and second order sum rules require the balance conditions between $I_\alpha^{\rm NS}(\omega)$ and $I_\alpha^{\rm SF}(\omega)$, and these conditions allow us to properly calculate spectra.

\subsubsection{Spectra of uncondensed normal state atoms}\label{sec_spec_uc}
%%%===%%%  t1
In what follows, we discuss the properties of $I_\alpha^{\rm NS}(\omega)$ and $I_\alpha^{\rm SF}(\omega)$ at finite temperatures,  separately.
Here we begin with $I_\alpha^{\rm NS}(\omega)$ by assuming that $N_{\rm SF}=0$, and accordingly $I_\alpha^{\rm SF}(\omega)$ vanishes.
We also assume that the inter-site correlations are negligible ${\cal G}_{i,j}$ for $i\not=j$ by comparing to the on-site correlations $n_i$.
The density matrix of such a localized state is given by $\prod_i (\sum_m e^{- E_{m,i}/T} |m\rangle_i\langle m|_i)$ at finite temperatures, where $E_{m,i}=U_{g,g}m(m-1)/2-V_{g,i}m$ is the energy of the local number state $|m\rangle_i$.
The spectra can be obtained in a form of the exact representation:
%%%===%%%  t2
\begin{eqnarray}
I_\alpha^{\rm NS}(\omega) & =& \sum_{i,m} w_{\alpha,i,m} \delta(\omega-p_{\alpha,i,m}). \label{eq_Leh}
\end{eqnarray}
The spectral weight $w_{\alpha,i,m}$ and the peak position $p_{\alpha,i,m}$ are given by
\begin{eqnarray}
w_{\alpha,i,m}&=& m {e}^{-(E_{m,i}-\Omega_i)/T}, \label{eq_w_NS} \\
p_{\alpha,i,m}&=&(m-1)(U_{g,e\alpha}-U_{g,g})+\Delta_{e\alpha}+V_{e\alpha,i}-V_{g,i}, \nonumber \\
\label{eq_p_NS}
\end{eqnarray}
where ${e}^{-(E_{m,i}-\Omega_i)/T}$ is the Boltzmann factor of the number state $|m\rangle_i$, and $\Omega_i(=-T \ln \sum_m e^{-E_{m,i}/T})$ is the grand potential in the $i$ th site.
Note that, even for the uniform systems, spectra at finite temperatures have multi-peak structures depending on the thermal distributions of the number states $|m\rangle_i$ described by ${e}^{-(E_{m,i}-\Omega_i)/T}$.

We now discuss that the zeroth and the first order sum rules are always satisfied in the above expression in Eq.\,(\ref{eq_Leh}) when ${\cal G}_{i,j} \ll n_i$.
By using Eqs.\,(\ref{eq_w_NS}) and (\ref{eq_p_NS}), we obtain $M_\alpha^{(0)} =\sum_{i,m}m{e}^{-\beta (E_{m,i}-\Omega_i)} $,
and $M_\alpha^{(1)} =\sum_{i,m}  m [(m-1)(U_{g,e\alpha}-U_{g,g})+\Delta_{e\alpha}+V_{e\alpha,i}-V_{g,i}]{e}^{-\beta (E_{m,i}-\Omega_i)}$.
Using relations $n_{i}=\sum_{m} m {e}^{-\beta (E_{m,i}-\Omega_i)}$ and $G_{2,i}=\sum_{m} m(m-1){e}^{-\beta (E_{m,i}-\Omega_i)}$,
 we find that $M_\alpha^{(0)}=N_{\rm NS}$, and $M_\alpha^{(1)}$ reduces to $\sum_i [(U_{g,e\alpha}-U_{g,g})G_{2,i} + (\Delta_{e\alpha}+V_{e\alpha,i}-V_{g,i})n_{i}]$.
These facts suggest that $I_\alpha^{\rm NS}(\omega)$ in Eq.\,(\ref{eq_Leh}) reproduces the zeroth and the first order moments in Eqs.\,(\ref{eq_mom_zth}) and (\ref{eq_mom_fst}) when we can neglect the term proportional to the inter-site correlations [the last term in Eq.\,(\ref{eq_mom_fst})].

In contrast, the second order sum rule is not straightforward.
When ${\cal G}_{i,j}\ll n_i$, almost all terms are reproduced in the same way as the above.
Namely, $M^{(2)}_\alpha= \sum_{i,m} w_{\alpha,i,m} p_{\alpha,i,m}^2$ is equivalent to the first four terms in Eq.\, (\ref{eq_mom_snd}).
However, we cannot reproduce one of the terms in Eq.\,(\ref{eq_mom_snd}), which is the round-trip hopping term given by $\sum_i \sum_{\bf d} |J_g-J_{e\alpha}e^{i {\bf k}_{\rm ex}\cdot {\bf d}}|^2 n_i $.
%As discussed in Sec.\ \ref{sec_discuss},
This means that, even though the inter-site correlations are negligible, quantum fluctuations resulting from the round-trip hopping broaden the spectral width of each peak in Eq.\,(\ref{eq_Leh}).
Namely, the sum rule requires that the delta function $\delta(\omega)$ in Eq.\,(\ref{eq_Leh}) should be replaced with a certain function with a finite spectral width.
We here use a Gaussian function, and $I_\alpha^{\rm NS}(\omega)$ is now given by
\begin{eqnarray}
&&I_\alpha^{\rm NS}(\omega)= \sum_{i,m}w_{\alpha,i,m}
\frac{\exp\left(- (\omega-p_{\alpha,i,m})^2/(2\gamma_\alpha^2)\right)}{\gamma_\alpha \sqrt{2\pi}},\nonumber \\ \label{eq_Leh_fw}
\end{eqnarray}
where $\gamma_\alpha$ is the spectral width defined by $\gamma_\alpha^2=\sum_{\bf d} |J_g-J_{e\alpha}e^{i {\bf k}_{\rm ex}\cdot {\bf d}}|^2$.
We comment that a Lorentzian function is not suitable for the substituting function, because the second order moment does not converge:
$\int^\infty_{-\infty} d\omega \omega^2 \gamma/\pi(\omega^2+\gamma^2) \to \infty$.

The extended representation in Eq.\,(\ref{eq_Leh_fw}) with Eqs.\,(\ref{eq_w_NS}) and (\ref{eq_p_NS}) properly satisfies the sum rules up to the second order when ${\cal G}_{i,j} \ll n_i$. %, when the inter-site correlations are negligible.
In the same way as the above, we can straightforwardly confirm that the zeroth and the first order sum rules are satisfied.
The second order $M_\alpha^{(2)}$ is extended as follows: $\int \omega^2 \sum_{i,m} w_{\alpha,i,m} e^{-(\omega-p_{\alpha,i,m})^2/2\gamma_\alpha^2}/(\gamma_\alpha \sqrt{2\pi})  d\omega=\sum_{i,m} w_{\alpha,i,m} p_{\alpha,i,m}^2+\gamma_\alpha^2 \sum_{i,m} w_{\alpha,i,m} $.
The first term $\sum_{i,m} w_{\alpha,i,m} p_{\alpha,i,m}^2$ is equivalent to the second order moment obtained from the original representation in Eq.\,(\ref{eq_Leh}).
The additional term $\gamma_\alpha^2 \sum_{i,m} w_{\alpha,i,m} (= \gamma_\alpha^2  N_{\rm NS})$ properly describes the last term in Eq.\,(\ref{eq_mom_snd}). %a finite spectral width resulting from the hopping.

We here estimate the magnitude of $\gamma_\alpha$ that characterizes an intrinsic spectral broadening caused by hopping-induced quantum fluctuations.
For simplicity, we consider $\alpha=1$ and set $J_g \sim J_{e1}$, which leads to $\gamma_1^2= 2zJ_{e1}^2 [1-\sum_{\bf d} \cos({\bf k}_{\rm ex}\cdot {\bf d})/z]$, where $z$ is the number of the neighboring lattice sites ($z=6$ in the cubic lattice).
For ${\bf k}_{\rm ex} \sim {\bf 0}$, $\gamma_1$ reduces to zero.
For ${\bf k}_{\rm ex} \sim (\pi,\pi,\pi)$, $\gamma_1$ takes a maximum $2\sqrt{z}|J_{e1}| = W_1/\sqrt{z}$, where $W_\alpha=2z|J_{e\alpha}|$ is a bandwidth of the $\alpha$ th orbital.
We next consider higher orbitals $\alpha \not = 1$ by assuming $|J_g| \ll |J_{e\alpha}|$, and then we obtain $\gamma_\alpha = \sqrt{z} |J_{e\alpha}|= W_\alpha/2\sqrt{z}$.
Simply put, the kinetic spectral broadening is proportional to the bandwidth, $\gamma_\alpha \propto W_\alpha$, where
the wavevector conservation law determines the proportionality coefficient ranging from 0 to $1/\sqrt{z}$ depending on ${\bf k}_{\rm ex}$.

Before closing the discussions on $I^{\rm NS}_\alpha(\omega)$, we consider the validity of the condition ${\cal G}_{i,j} \ll n_i$.
%, which assures that the above expression is appropriate.
%when the inter-site correlations ${\cal G}_{i,j}$ is sufficiently smaller than the on-site ones $n_i$ .
For $J_{g}=0$, this condition is exactly satisfied: ${\cal G}_{i,j}=0$ for $i\not=j$.
For a finite but small $J_g (\ll U_{g,g})$, where MI states will appear at low temperatures,  %will be small because of the many-body co
the effects of interactions strongly suppress the inter-site correlations.
Large potential differences strongly suppress the inter-site correlations (e.g., $|V_{g,i}- V_{g,j}| \gg J_{g}$), and thermal fluctuations also decrease ${\cal G}_{i,j}$. % when temperature $T$ is comparable to bandwidth $W_{\alpha=1}$.
We thus expect that the NS atoms in the realistic systems with interactions and trapping potential at finite temperatures will satisfy well the condition of small ${\cal G}_{i,j} (\ll n_i)$. %\langle \hat{c}_{g,i}^\dag \hat{c}_{g,j}\rangle$.
This condition is equivalent to flattened momentum distributions $n_{g, \bf k}$, which can be confirmed in experiments by using the time-of-flight measurements with the projection onto the first Brillouin zone \cite{BandMapping}.

\subsubsection{Spectra of the superfluid atoms}

Next,  we consider the opposite limit, $N_{\rm SF} \gg N_{\rm NS}$, where we neglect $I_\alpha^{\rm NS}(\omega)$.
Taking account of the physical properties of BEC,
%$I_\alpha^{\rm SF}(\omega)$ mentioned in Sec.\ \ref{sec_discuss},
we assume that $I_\alpha^{\rm SF}(\omega)$ can be described by the following single peak structure:
\begin{eqnarray}
I_\alpha^{\rm SF}(\omega)&=& N_{\rm SF} \frac{\exp\left(-(\omega-\bar{\omega}_\alpha^{\rm SF})^2/2(\sigma_\alpha^{\rm SF})^2 \right)}{(\sigma_\alpha^{\rm SF}\sqrt{2\pi})},\nonumber\\
\label{eq_spcSF}
\end{eqnarray}
where a spectral peak position $\bar{\omega}_\alpha^{\rm  SF}$ and a spectral width $\sigma_\alpha^{\rm SF}$ are determined from the sum rules in Eqs.\,(\ref{eq_BFsum_snd}) and (\ref{eq_BFsum_trd}), respectively.
For such a single peak structure, the deviation coincides with the spectral width.
This single peak assumption may be oversimplification. %is  $I_\alpha^{\rm SF}(\omega)$ has a single peak only,  which
It should be noted that we carefully take account of the spectral broadening caused by number fluctuations, which allows us to reasonably use this simple assumption.

To compare $I^{\rm SF}_\alpha(\omega)$ with $I^{\rm NS}_\alpha(\omega)$, here we mention again the properties of $I^{\rm NS}_\alpha(\omega)$ at finite temperatures.
As shown in Eq.\,(\ref{eq_Leh_fw}), $I_\alpha^{\rm NS}(\omega)$ has many peaks at positions $p_{\alpha,i,m} (\propto U_{g,e\alpha}-U_{g,g})$  with a spectral width of $\gamma_\alpha (\propto W_\alpha)$. % |J_{g}-J_{e\alpha}e^{i{\bf k}_{\rm ex}\cdot {\bf d}}|
Note that peak positions and a width are usually determined from the different energy scales.
Thermal fluctuations change relative spectral weights and also increase the number of spectral peaks. %, which is properly taken abbount of in this formulation.
This multi-peak structure is an essential feature of $I_\alpha^{\rm NS}(\omega)$ at finite temperatures, which can be properly dealt with in Eq.\,(\ref{eq_Leh_fw}).
In contrast, as discussed in Sec.\ \ref{sec_discuss}, for SF atoms, a spectral mean value is proportional to the average number of atoms $\bar{\omega}_\alpha^{\rm  SF}\propto (U_{g,e\alpha}-U_{g,g})\bar{n}$, and a spectral deviation is  given by $\sigma_\alpha^{\rm SF}\sim |U_{g,e\alpha}-U_{g,g}|\sqrt{\bar{n}}$.   %the averaged number of atoms $\bar{n}$  as connected to number fluctuations
This fact suggests that $\bar{\omega}_\alpha^{\rm  SF}$ and $\sigma_\alpha^{\rm SF}$ will be usually comparable.
A main role of thermal fluctuations is a decrease in the number of condensed atoms $N_{\rm SF}$.
%, while other properties of BEC will be hardly changed when only temperature changes.
%:  As temperature changes, the lowest energy level $E_g$ of the Hamiltonian ${\cal H}_g$ is unchanged, while the distributions of atoms change.
We thus expect that the essence of $I_\alpha^{\rm SF}(\omega)$ at finite temperatures can be captured by a single peak broadened by large number fluctuations.
The effects of decrease in $N_{\rm SF}$ on the spectra are considered via calculations on the sum rules in Eq.\,(\ref{eq_BFsum_fst}), and also Eqs.\,(\ref{eq_BFsum_snd}) and (\ref{eq_BFsum_trd}), which properly describe decreases in a peak height, a peak shift and broadening, respectively.
 %, which is a key to the properties of the SF-MI transition.
%Note that, when we consider the coexisting states,  Eqs.\,(\ref{eq_BFsum_snd}) and (\ref{eq_BFsum_trd}) include
%not only the effects of number fluctuations
%Note that a spectral broadening determined from Eq.\,(\ref{eq_BFsum_trd}) is affected by not only large number fluctuations but also kinetic contributions, because the coexisting states of SF and NS atoms have finite number and also phase fluctuations.
%In Eqs.\,(\ref{eq_BFsum_snd}) and (\ref{eq_BFsum_trd}), correlations between SF and NS atoms are also included effectively.
%The validity of our assumptions will be checked in Sec. \ref{sec_results} by comparing the calclated spectra and those in experiments.

\subsubsection{Spectra for coexisting region}
We next explain the formulation for the middle region, $N_{\rm SF}\not=0$ and $N_{\rm NS}\not=0$, where NS and SF atoms coexist.
Here, $I_\alpha^{\rm SF}(\omega)$ and $I_\alpha^{\rm NS}(\omega)$ are separately calculated, and
$I_\alpha^{\rm NS}(\omega)$ is the first.
We assume that NS atoms are affected by a mean-field potential resulting from interactions with SF atoms.
Thus, we here consider the following effective local Hamiltonian of NS atoms excluding SF atoms:
\begin{eqnarray}
\hat{\cal H}_{{\rm NS}}&=& %\sum_{j}J_{g}\hat{c}_{{\rm NS},{i}}^\dag\hat{c}_{{\rm NS},{j}}+
\sum_i (V_{g,i}+V_{{\rm SF},i}) \hat{n}_{{\rm NS},i} \nonumber \\
&+&U_{g,g}\sum_i \hat{n}_{{\rm NS},i}(\hat{n}_{{\rm NS},i}-1)/2, \label{eq_effHmi}
\end{eqnarray}
where the potential $V_{{\rm SF},i}$ describes effectively mean-field interactions between SF and NS atoms given by $V_{{\rm SF},i}=2U_{g,g} n_{{\rm SF},i}+\delta\mu_i$.
To impose a self-consistent condition $\langle \hat{n}_{{\rm NS},i}\rangle_{\hat{\cal H}_{{\rm NS}}}=\langle \hat{n}_{g,i}\rangle-|\langle \hat{c}_{g,i}\rangle|^2$, we further define a chemical potential shift $\delta\mu_i$, where $\langle \cdots\rangle_{\hat{\cal H}_{{\rm NS}}}$ is  the statistical average at thermal equilibrium defined by the Hamiltonian $\hat{\cal H}_{{\rm NS}}$, while $\langle \cdots\rangle$ is that defined by the localized Hubbard Hamiltonian in Eq.\,(\ref{eq_HmiLoc}).
We note that $\delta\mu_i \sim 0$ for $n_{{\rm SF},i} \gg n_{{\rm NS},i}$ or $n_{{\rm SF},i} \ll n_{{\rm NS},i}$, because the mean-field treatment is appropriate for these dilute regions.

We summarize a procedure for calculating the full spectra $I(\omega)=I^{\rm SF}(\omega)+I^{\rm NS}(\omega)$, where
$I^{\rm SF}(\omega)\equiv \sum_\alpha |\rho_\alpha|^2 I_\alpha^{\rm SF}(\omega)$ and $I^{\rm NS}(\omega)\equiv \sum_\alpha |\rho_\alpha|^2 I_\alpha^{\rm NS}(\omega)$.
\begin{enumerate}
\item We first calculate thermal equilibrium states of Hubbard Hamiltonian based on the finite temperature Gutzwiller approximation.
We use exact diagonalization to solve the localized Hamiltonian in Eq.\,(\ref{eq_HmiLoc}) at finite temperatures.
We obtain the moments $M_\alpha^{(n)}$ from Eqs.\,(\ref{eq_mom_zth})-(\ref{eq_mom_snd}), and other statistical quantities such as $n_{{\rm SF},i}$ and $n_{{\rm NS},i}$.
\item Next, we calculate $I_\alpha^{\rm NS}(\omega)$ in Eq.\,(\ref{eq_Leh_fw}) by exactly diagonalizing the effective Hamiltonian in Eq.\,(\ref{eq_effHmi}).
After that, we can directly calculate $\bar{\omega}_\alpha^{\rm NS}$ and ${\sigma}_\alpha^{\rm NS}$ from $I_\alpha^{\rm NS}(\omega)$.
For the self-consistent condition mentioned above, we need $n_i$ and ${c}_{i}$, which should be obtained in the previous process 1.
\item Finally, we determine $\bar{\omega}_\alpha^{\rm SF}$ and $\sigma_\alpha^{\rm SF}$ by using Eqs.\,(\ref{eq_BFsum_snd}) and (\ref{eq_BFsum_trd}), where we use $M_\alpha^{(n)}$, $\bar{\omega}_\alpha^{\rm NS}$, and ${\sigma}_\alpha^{\rm NS}$ obtained in the previous processes 1 and 2. Then, we can calculate the full spectra $I(\omega)$.
\end{enumerate}
In this way, based on the sum rule approach, completely different features of $I_\alpha^{\rm NS}(\omega)$ and $I_\alpha^{\rm SF}(\omega)$ are properly dealt with, and multi-peak structures resulting from finite temperature effects and coexisting of SF and NS atoms will be taken account of precisely.

\subsection{Role of the sum rules in the two-mode approximation}

We finally discuss what a role the sum rules play in the present calculation procedure.
The sum-rule approach combined with the two-mode approximation provides us with the
reasonable relationship between $I_\alpha^{\rm NS}(\omega)$ and $I_\alpha^{\rm SF}(\omega)$.
As discussed below, $\bar{\omega}_\alpha^{\rm SF}$ and ${\sigma}_\alpha^{\rm SF}$ can be determined reasonably within the level of the mean-field treatment.

We here consider the first order moment $M^{(1)}_\alpha (= N_{\rm NS}\bar{\omega}_{{\rm NS},\alpha}+N_{\rm SF}\bar{\omega}_{{\rm SF},\alpha})$ and focus on the effects of interactions [the first term in Eq.\,(\ref{eq_mom_fst})]: $(U_{g,e\alpha}-U_{g,g})\sum_i G_{2,i}$, and other terms are neglected for clarity.
Using a definition of $\hat{c}_{NS,i}=\hat{c}_{g,i}-c_i$, we can rewrite $G_{2,i}$ as $G^{\rm NS}_{2,i}+ 4 n_{{\rm NS},i} n_{{\rm SF},i}+ n_{{\rm SF},i}^2 $, where $G^{\rm NS}_{2,i}\equiv \langle \hat{c}_{{\rm NS},i}^\dag \hat{c}_{{\rm NS},i}^\dag  \hat{c}_{{\rm NS},i} \hat{c}_{{\rm NS},i} \rangle$.
Three terms, $G^{\rm NS}_{2,i}$, $n_{{\rm NS},i} n_{{\rm SF},i}$, and $n_{{\rm SF},i}^2$, represent the two-body correlations between two NS atoms, between NS and SF atoms, and between two SF atoms, respectively.
%\equiv \langle \hat{c}_{g,i}^\dag \hat{c}_{g,i}^\dag  \hat{c}_{g,i} \hat{c}_{g,i} \rangle$
As discussed in Sec.\ \ref{sec_spec_uc}, the first order moment of $I^{\rm NS}_\alpha(\omega)$ is easily obtained:
$$N_{\rm NS}\bar{\omega}^{\rm NS}_{\alpha}=(U_{g,e\alpha}-U_{g,g}) \sum_i  (G^{\rm NS}_{2,i}+ 2 n_{{\rm NS},i} n_{{\rm SF},i}),$$ where the latter term results from the mean field potential $V_{{\rm SF},i}$ in Eq.\,(\ref{eq_effHmi}).
Consequently, the sum rule in Eq.\,(\ref{eq_BFsum_snd}) allows us to determine the correlation term in the first moment of $I_\alpha^{\rm SF}(\omega)$: $$N_{\rm SF}\bar{\omega}^{\rm SF}_{\alpha}= (U_{g,e\alpha}-U_{g,g}) \sum_i (n^2_{{\rm SF},i} +2 n_{{\rm SF},i}n_{{\rm NS},i}).$$
The SF-NS correlation $4 n_{{\rm NS},i} n_{{\rm SF},i}$  is shared equally between $I_\alpha^{\rm NS}(\omega)$ and $I_\alpha^{\rm SF}(\omega)$.
We now again consider the uniform system to compare this expression with those in Sec.\ \ref{sec_discuss}.
%Within the local approximation, we can easily calculate the kinetic energy shift of the SF atoms.
The collisonal energy shift for the SF atoms is now given by $\bar{\omega}^{\rm SF}_{\alpha}=(U_{g,e\alpha}-U_{g,g})(n_{{\rm SF},i} +2 n_{{\rm NS},i})$. %(\varepsilon_{e\alpha,{{\bf k}_0+{\bf k}}_{\rm ex}}-\varepsilon_{g,{\bf k}_0})
The first term $(U_{g,e\alpha}-U_{g,g})n_{{\rm SF},i}$ is equivalent to that of the pure SF atoms as discussed in Sec.\ \ref{sec_discuss} and also in the previous study \cite{Hazzard}, while the second term is the additional contribution originating from the mean-field NS-SF interactions.
%Kinetic energy shift is the same as that of the pure SF atoms.
%provides a reasonable extension of a method for calculating spectra in the coexisting region from zero to finite temperatures

In the same way as the above, we can obtain the correlation terms in the second order moment of $I_\alpha^{\rm NS}(\omega)$ and $I_\alpha^{\rm SF}(\omega)$: $N_{\rm NS}[({\sigma}^{\rm NS}_{\alpha})^2 +(\bar{\omega}^{\rm NS}_{\alpha})^2]=(U_{g,e\alpha}-U_{g,g})^2 \sum_i [G^{\rm NS}_{3,i}+G^{\rm NS}_{2,i}(1+4n_{{\rm SF},i})+4n_{{\rm NS},i}n_{{\rm SF},i}^2]$,
and $N_{\rm SF}[({\sigma}^{\rm SF}_{\alpha})^2 +(\bar{\omega}^{\rm SF}_{\alpha})^2] = (U_{g,e\alpha}-U_{g,g})^2 \sum_i [n^3_{{\rm SF},i}+n^2_{{\rm SF},i}(1+5n_{{\rm NS},i}) +n_{{\rm SF},i}(4n_{{\rm NS},i}^2+5G^{\rm NS}_{2,i})]$.
For the uniform system, $({\sigma}^{\rm SF}_{\alpha})^2$ is written as $(U_{g,e\alpha}-U_{g,g})^2 [n_{{\rm SF},i}+n_{{\rm SF},i} n_{{\rm NS},i} +G^{\rm NS}_{2,i}+4(\Delta n_{{\rm NS},i})^2] $, where $(\Delta n_{{\rm NS},i})^2= \langle \hat{n}_{{\rm NS},i}^2 \rangle -n_{{\rm NS},i}^2$.
Note that $\Delta n_{{\rm NS},i} \sim 0$ at low temperatures.
The first term corresponds to number fluctuations of the SF atoms, and the other terms suggest that the NS-SF correlations enhance number fluctuations of the SF atoms and further broaden $I_\alpha^{\rm SF}(\omega)$.
%We skip to discuss other terms such as kinetic and trapping potential terms.
%Instead, we perform numerical

%, where $(\Delta n_{{\rm NS},i})^2= \langle \hat{n}_{{\rm NS},i}^2 \rangle -n_{{\rm NS},i}^2$
%$N_{\rm NS}\bar{\omega}_{{\rm NS},\alpha}$ and $N_{\rm SF}\bar{\omega}_{{\rm SF},\alpha}$.
%Kinetic energy shift $\delta K_\alpha$ is straightforwardly obtained as $\bar{\omega}_{\rm SF}=\sum_{\bf k} (\varepsilon_{e\alpha,{\bf k+k}_{\rm ex}}-\varepsilon_{g,{\bf k}}) n_{g,{\bf k}}/N_{\rm tot}$
%We should note that a two-mode approximation is
%In the same way as the above, the condition in Eq.\,(\ref{eq_BFsum_trd}) allows us to effectively discuss the SF-NS correlation effects on the spectral deviations within the mean field approximation.

\subsection{Comparison with the previous studies}\label{sec_comp}
%%%===%%%  t3
Here, to discuss difference between the present and the previous treatment, we summarize the previous formulation \cite{Hazzard}, which has been successfully applied to the microwave spectroscopy experiments \cite{Campbell}:
\begin{equation}
I_{}(\omega)= \sum_{i} n_{i} \delta\left(\omega - (U_{g,e1}-U_{g,g}) G_{2,i}/n_{i}\right).  \label{eq_Haz}
\end{equation}
This approximation satisfies the zeroth order sum rule $M^{(0)}_\alpha=N_{\rm tot}$ and partly satisfies the first order, $M^{(1)}_\alpha=\sum_i G_{2,i} (U_{g,e\alpha}-U_{g,g})$, while the second order sum rule is not satisfied at all.
Note that, without a consideration of the deviation $\sigma_\alpha$, the spectral broadening is effectively described by the site-dependent $G_{2,i}$ and $n_i$ resulting from the inhomogeneity of the system.
When $k_{\rm ex}=0$,  $|J_g-J_{e\alpha} e^{i{\bf k}_{\rm ex}\cdot {\bf d}}|=0$, $V_{e\alpha,i}-V_{g,i}=0$,  $T=0$ and the coexisting of NS and SF states is neglected (either $N_{\rm NS}=0$ or $N_{\rm SF}=0$), the spectral position $\bar{\omega}_\alpha^{\rm SF}$ obtained from Eq.\,(\ref{eq_mom_fst}) or ${p}_{i,\alpha,m}$ in Eq.\,(\ref{eq_p_NS}) and the corresponding term $(U_{g,e1}-U_{g,g}) G_{2,i}/n_{i}$ in Eq.\,(\ref{eq_Haz}) are equivalent with each other.
Thus, within the first order approximation, our method is consistent with Eq.\,(\ref{eq_Haz}) in the following two limits; pure SF or MI phases at zero temperature.
We note that, at least, the conditions $k_{\rm ex}=0$,  $|J_g-J_{e\alpha} e^{i{\bf k}_{\rm ex}\cdot {\bf d}}|=0$, and $V_{e\alpha,i}-V_{g,i}=0$ are well satisfied in the microwave spectroscopy (see Sec.\ \ref{sec_param}).

%The approximation described by Eq.\,(\ref{eq_Haz}) assumes that spectra consist of a single peak structure in each site.
%However, at finite (higher) temperatures, multi-peak structures are required as discussed above, and even at zero temperature, the coexisting of SF and NS atoms will occur due to the correlation effects.
%Thus, Ref. \cite{Hazzard2010} provides the extended two-peak formulation, and our method can describe the multi-peak structures at finite temperatures.

\section{Numerical simulations}\label{sec_results}
In this section, we show numerical results calculated  by considering realistic parameters of the following two experiments; the microwave-spectroscopy of $^{87}$Rb atoms and the laser spectroscopy of $^{174}$Yb atoms.
We first explain the parameters and then discuss obtained results.

\subsection{Parameters}\label{sec_param}
%%%===%%%  r1
We first point out intrinsic differences of the microwave and the laser spectroscopy, and provide parameters used in the calculations.
A length of microwave is much longer than a lattice constant $a_L(=\lambda_L/2)$ and a lattice laser wavelength $\lambda_L$ (e.g., of $1064$ nm), and as a result, a wavevector $k_{\rm ex}$ can be set zero.
Therefore, the parameter region of the microwave spectroscopy corresponds to the perfect Lamb-Dicke regime, where excitation matrices $\rho_\alpha$ in Eq.\,(\ref{eq_rhoalpha}) are given by $\rho_1=1$ and $\rho_{\alpha\not=1}=0$.
Here the kinetic energy shift $\delta{K}_\alpha$ and deviations $\sigma_{K, \alpha}$ and $\sigma_{\sqrt{UK}, \alpha}$ are also negligible.
In contrast, for the laser spectroscopy, a wavelength of the excitation laser $\lambda_{\rm ex}$ (e.g., of 507 nm) is comparable to $\lambda_L$ (e.g., of 532 nm).
Namely, $k_{\rm ex}$ is the same order as a lattice wavevector $2\pi/a_L$.
Thus, orbital-changing excitations and kinetic contributions will be important in the laser spectroscopy.

The microwave spectroscopy of $^{87}$Rb atoms uses an excitation between different hyperfine states \cite{Campbell}, while
the laser spectroscopy of $^{174}$Yb atoms uses an excitation between different electron configurations, $^1$S$_0$ and $^3$P$_2$ states \cite{NJP:507,3P2MRI:App,YamashitKato}. % $^1$S$_0$ to $^3$P$_2$ states
Rb atoms are trapped by the combination of optical and magnetic potential, and Yb atoms are trapped with optical potential.
For the microwave spectroscopy, we use a harmonic trapping potential $V_{g,i} \propto C_{x} x_i^2+C_{y} y_i^2+C_{z} z_i^2$, where the curvatures $C_{x}$, $C_{y}$, and $C_{z}$ are determined from the experimental parameters so as to reproduce the bottom of trapping potential \cite{Campbell}, while for the laser spectroscopy, we use an anharmonic potential by carefully considering the laser configurations in experiments \cite{ExpPaper}.
Note that the trapping potential of two hyperfine states was set to be the nearly same \cite{Campbell}, while $^1$S$_0$ and $^3$P$_2$ states of Yb atoms are trapped in the different potential due to the greatly different polarizability. %Stark coefficients.
Thus, $V_{e\alpha,i}-V_{g,i}$ can be set zero for the microwave spectroscopy, while it is finite for the laser spectroscopy.

Differences in scattering lengths $a_{g,e}-a_{g,g}$ are  -0.13 nm and -30 nm for the microwave and laser spectroscopy, respectively.
Both of them are negative, so that differences in the interaction strengths $U_{g,e\alpha}-U_{g,g}$ are also negative.
In addition to the spectral broadening caused by quantum fluctuations (see Sec.\ \ref{sec_discuss}), we consider linewidths of the excitation laser of about 1 kHz and of the microwave of about 5 Hz, including the Fourier width of the excitation pulse.

\subsection{Comparison between two spectroscopy for the deep lattice}\label{sec_rf_laser}
\begin{figure}[t]
\includegraphics[width=9cm]{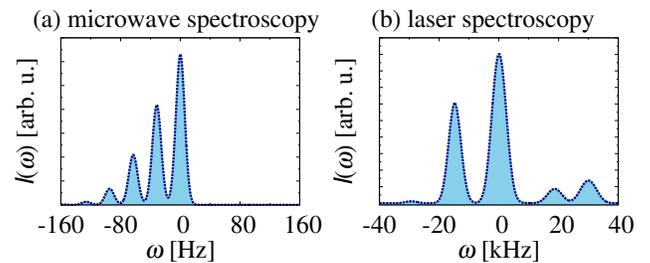}
\caption{(Color Online) Spectra $I(\omega)$ for the microwave and laser spectroscopy in a deep lattice calculated by the following parameters:
(a) $V_0=35 E_r$, $N_{\rm tot}=10^5$, and $T=100$ nK, and
(b) $V_0=15 E_r$, $N_{\rm tot}=2.2\times 10^4$, and $T=100$ nK.
There are no SF atoms because of the strong interactions; $N_{\rm SF}\sim 0$ and $I(\omega)\sim I^{\rm NS}(\omega)$.
}
\label{fig_both}
\end{figure}

%%%===%%%  r1
We first discuss spectra in a deep lattice and compare two kinds of spectroscopy.
Figure\ \ref{fig_both} shows the spectra calculated at a temperature $T$ of $100$ nK.
The other parameters in the microwave spectroscopy are $V_0=35 E_r$ and $N_{\rm tot}=10^5$, and those in the laser spectroscopy are $V_0=15 E_r$ and $N_{\rm tot}=2.2\times 10^4$, where $E_r$ is the recoil energy.
A spectral peak appearing at $\omega=0$ always means that the $m=1$ number state ($|m=1\rangle$) is excited without orbital changing, because the origin of spectra is renormalized by setting $\Delta_{e1}=0$.
Since $U_{g,e\alpha}-U_{g,g}$ is negative for both spectroscopy, peaks of $|m\rangle$ with $m \ge 2$ appear orderly in the region of $\omega <0$.
The orbital-changing excitation requires a large positive bandgap energy $\Delta_{e2}$.
Thus, spectra in the laser spectroscopy show some peaks in $\omega > 0$, which have the similar characteristics to those in $\omega<0$ but have the small intensities because of a small excitation probability $|\rho_{2}|^2 (\sim 0.1 |\rho_{1}|^2)$. % and $\rho_{2,1,1}$.

Next, we discuss a spectral peak width. %, which may be broadened by certain many-body effects.
For a deep lattice, spectral broadening is mainly attributed to the effects of inhomogeneity.
Equation\,(\ref{eq_mom_fst}) shows that potential energy difference $(V_{e\alpha,i}-V_{g,i})n_i/N_{\rm tot}$ causes just a peak shift.
However, due to the inhomogeneity, a variation of $(V_{e\alpha,i}-V_{g,i})n_i/N_{\rm tot}$ for different $i$ effectively induces broadening of spectra.
For the laser spectroscopy, an energy scale of this variation is estimated to be about 1 kHz, which is consistent with the obtained spectral features in Fig.\ \ref{fig_both}.
In the present parameter region, broadening effects resulting from the hopping terms are negligible by comparing with this inhomogeneous broadening.
On the other hand, since $|V_{e\alpha,i}-V_{g,i}| \sim 0$ for the present microwave spectroscopy, a width of each peak nearly equals to a linewidth of the microwave pulse. % of about $5$ Hz.

\subsection{Lattice depth dependence of spectra in the microwave spectroscopy}\label{sec_vdep}
%%%===%%%  r1
\begin{figure}[t]
\includegraphics[width=5cm]{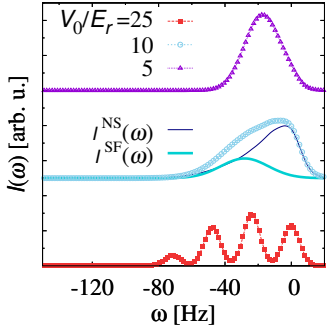}
\caption{(Color Online) Spectra $I(\omega)$ of the microwave spectroscopy for $N_{\rm tot}=10^5$ at $T=25$ nK for $V_0=25 E_r$, $10 E_r$ and $5 E_r$.
For $V_0=10 E_r$, the contributions of SF and NS atoms, $I^{\rm SF}(\omega)$ and $I^{\rm NS}(\omega)$, are also shown by thick and thin lines, respectively, where $I(\omega)=I^{\rm SF}(\omega)+I^{\rm NS}(\omega)$.
For $5 E_r$, $N_{\rm NS} \sim 0$ and $I(\omega)\sim I^{\rm SF}(\omega)$, while
for $25 E_r$, $N_{\rm SF} \sim 0$ and $I(\omega)\sim I^{\rm NS}(\omega)$.
}\label{fig_vdep}
\end{figure}

Next,  by focusing on the microwave spectroscopy, we discuss how spectra change as the lattice depth varies.
Figure\ \ref{fig_vdep} shows the spectra $I(\omega)$ calculated with $V_0=5 E_r, 10 E_r,$ and $25 E_r$ for $N_{\rm tot}=10^5$ at $T= 25$ nK.
For $V_0=25 E_r$, the number of SF atoms $N_{\rm SF}$ is zero, and  a discrete peak structure resulting from $I^{\rm NS}(\omega)$ appears.
In contrast, for $V_0=5 E_r$, almost all atoms are in condensed states $N_{\rm SF}\sim N_{\rm tot}$ and $N_{\rm NS}\sim 0$.
From discussions in Sec.\ \ref{sec_sumrule}, we can naively expect that the spectra of SF atoms $I^{\rm SF}(\omega)$ are centered at around $(U_{g,e\alpha}-U_{g,g})\bar{n}$, and a peak width is determined from $|U_{g,e\alpha}-U_{g,g}|\sqrt{\bar{n}}$, where $\bar{n}$ is the averaged number of atoms. % and is broadened by number fluctuations.
From these features, we can estimate $\bar{n} \sim 2$.
For a middle region $V_0=10 E_r$, we find an asymmetric broadened peak structure.
We note that this characteristic asymmetric structure can be attributed to the coexisting of NS and SF atoms.
Figure\ \ref{fig_vdep} also shows spectra of SF atoms $I^{\rm SF}(\omega)$ and those of NS atoms $I^{\rm NS}(\omega)$.
The spectra of SF atoms $I^{\rm SF}(\omega)$ are centered at around $2.5(U_{g,e\alpha}-U_{g,g})$.
On the other hand, $I^{\rm NS}(\omega)$ has a large intensity at around $\omega \sim 0$ corresponding to the $|m=1\rangle$ excitation.
Here we find no discrete peak structures, because $|U_{g,e\alpha}-U_{g,g}|$ is smaller than the microwave linewidth.
The sum of two spectra yields the characteristic asymmetric spectra.

The obtained spectra for all three parameters in Fig.\ \ref{fig_vdep} capture essential features of those observed in experiments \cite{Campbell}:  The experimental spectra show the symmetric single peak structure for $V_0 =5 E_r$, and the asymmetric broadened peak structure, or an overlapped double-peak structure, for  $V_0 =10 E_r$, and the discrete peak structure for $V_0=25 E_r$.
Our assumptions, {\it e.g.,} the single peak assumption for the SF atoms, and the two-mode approximation, appropriately reproduce characteristic structures of the spectra seen in experiments.

%%%===%%%  r1
\subsection{Temperature dependence of spectra in the microwave spectroscopy}\label{sec_tdep}
\begin{figure}[t]
\includegraphics[width=6cm]{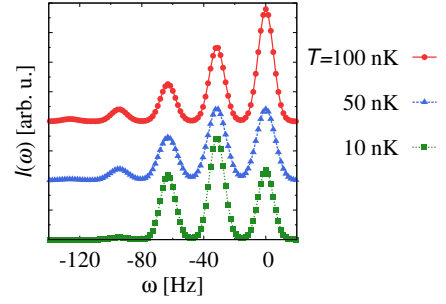}
\caption{(Color Online) Spectra $I(\omega)$ for the microwave spectroscopy for $V_0=35 E_r$ and $N_{\rm tot}=10^5$ at $T=$ 100 nK, 50 nK, and 10 nK.
}
\label{fig_tmpDep}
\end{figure}

\begin{figure}[t]
\includegraphics[width=9cm]{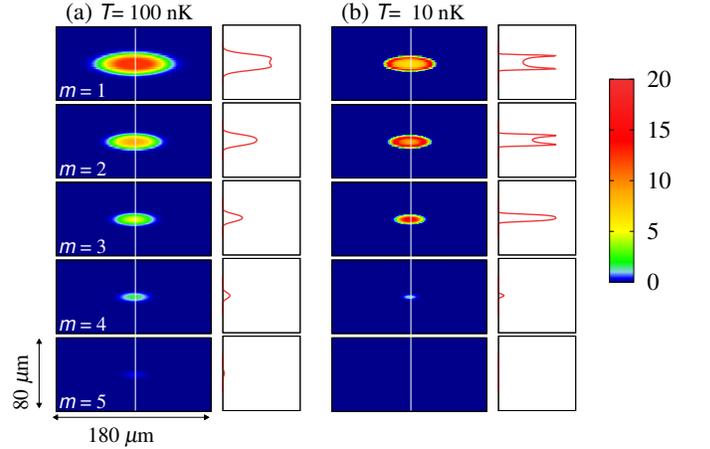}
\caption{(Color Online) In-situ distribution of the number states $|m\rangle$  for $V_0=35 E_r$ and $N_{\rm tot}=10^5$ at $T$= 100 nK (a) and 10 nK (b), and cross sections along the (white) line. % at the center.
}\label{fig_num}
\end{figure}

Next, we discuss the temperature dependence of spectra by focusing on the microwave spectroscopy again.
In Fig.\ \ref{fig_tmpDep}, we first show the spectra for $V_0=35 E_r$ at $T= 100$ nK, 50 nK, and 10 nK.
From Fig.\ \ref{fig_tmpDep}, we find that spectral peak height of each $|m\rangle$ state varies depending on temperature, which indicates the change in thermal distributions of the number states $|m\rangle$.
At low temperatures ($T=$10 nK),  the $m=2$ peak is highest.
On the other hand, at higher temperatures ($T=$100 nK and 50 nK), the $m=1$ peak is dominant.
To clearly discuss this behavior, in Fig.\ \ref{fig_num}, we show the in-situ column-density distributions of the number states $|m\rangle$, which can be calculated by exact diagonalization for the localized Hamiltonian in Eq.\,(\ref{eq_HmiLoc}).
At higher temperatures ($T=$100 nK),  the $|m=1\rangle$ state spreads widely, while larger-$m$ states prefer center of the potential.
Thus, $N_m$ the total number of $|m\rangle$ monotonically decreases with increasing $m$.
This behavior is consistent with those seen in spectra in Fig.\ \ref{fig_tmpDep}.
In contrast, at low temperatures ($T=$10 nK), the Mott shell structures develop, and as a result, the number state distributions for $m \le2$ show a dip structure as shown in Fig.\ \ref{fig_num}.
Here a three dimensional shell structure is projected onto the dip structure in the column-density distributions.
Thus, $N_m$ becomes nonmonotonic and $N_{m=2}$ becomes maximum, leading to a large intensity of the  peak of $|m=2\rangle$ in spectra (see Fig.\ \ref{fig_tmpDep}).
These features suggest that the present spectroscopy can be a thermometer.

We finally compare these results with the experimental observations  in Ref.\ \cite{Campbell}, in which spectra and also in-situ number-state distributions have been measured.
We find that both spectra and distributions agree well with the calculations at higher temperatures than those at lower temperatures.
It thus suggests that temperature will be an important parameter to discuss the quantitative aspect of experiments.
Although nearly pure BEC can be created before loading atoms into a lattice,  the loading usually induces a certain amount of heating  \cite{2011:Sugawa:DualMott}.
%parameters are not precisely taken account of, although we set trapping frequencies at around the bottom to be the same as the experimental parameters \cite{Campbell}.
In the present calculations, a harmonic trapping potential is used by focusing on the bottom of the experimental trapping potential \cite{Campbell}, whereas in Ref. \cite{Hazzard}, by carefully considering the anharmonicity of trapping potential, experimental spectra can be reproduced even at zero temperature.
The calculations considering both anharmonic potentials and thermal fluctuations may be required for more quantitative agreements.

\section{Summary}
In summary, we theoretically investigate the microwave and the laser spectroscopy on the bosonic Hubbard systems.
We first discuss the sum rules of spectra up to the second order, which can be derived from the general principle of spectroscopy.
This principle provides various useful information on the physical properties of spectra.
The spectra of superfluid states with phase coherence is broadened by the many-body effects, and its broadened width can be characterized by number fluctuations in thermal equilibrium.
In contrast, the spectra of the number definite Mott insulating states are broadened by quantum fluctuations caused by tunneling effects. %, which reflects the uncertainty principle.

We next propose a two-mode approximation to calculate spectra at finite temperatures.
This approximation assumes that spectra can be decomposed into two contributions originated from Bose-Einstein condensates and uncondensed normal states.
Our method is built up by considering the spectral characteristics figured out from the sum rules, so that the multi-peak structures resulting from coexisting of superfluid and normal states at finite temperatures can be successfully dealt with.

Finally, by combining the two-mode approximation with the finite temperature Gutzwiller approximation, we numerically calculate spectra by considering realistic experimental parameters of the microwave and the laser spectroscopy.
We find that our method can reproduce the essential features of spectra in experiments.
We also discuss the lattice depth dependence and the temperature dependence of the microwave spectra.
These results clarify that the present spectroscopy can be sensitive tools for investigating the quantum phase transitions at finite temperatures.

\begin{acknowledgments}
We acknowledge Y. Takahashi and S. Kato for discussing their experiments.
\end{acknowledgments}


\begin{references}
\newcounter{conf}
\setcounter{conf}{72}
%%%%%%%%%%%%%%%%%%%%%%%%%%%%%%%%%%%%%%%%%%%%%%%%%%%%%%%%
%Reviews
\bibitem{Review_Jaksch}
     D. Jaksch and P. Zoller,
     Ann. Phys. {\bf 315}, 5279 (2005).
\bibitem{Review_Bloch}
     I. Bloch, J. Dalibard and W. Zwerger,
     Rev. Mod. Phys. {\bf 80}, 885 (2008).
%%%%%%%%%%%%%%%%%%%%%%%%%%%%%%%%%%%%%%%%%%%%%%%%%%%%%%%%
%Experiments
%%%Interfrerence
\bibitem{SFtoMott_Greiner:Nat}
     M. Greiner, O. Mandel, T. Esslinger, T. W. H\"{a}nsch, and I. Bloch,
     Nature {\bf 415}, 39 (2002).
\bibitem{Phase_Gerbier:PRL}
     F. Gerbier, A. Widera, S. F\"{o}lling, O. Mandel, T. Gericke, and I. Bloch,
     Phys. Rev. Lett. {\bf 95}, 050404 (2005).
	\bibitem{YbMott:PRA}
     T. Fukuhara, S. Sugawa, M. Sugimoto, S. Taie, and Y. Takahashi,
     Phys. Rev. A {\bf 79}, 041604R (2009).
%\bibitem{2DMott:PRL}
%     I. B. Spielman, W. D. Phillips, and J. V. Porto,
%     Phys. Rev. Lett. {\bf 98}, 080404 (2007).
%%%Noise Correlation
%%%
	\bibitem{NoiseCorr:Nat}
     S. F\"{o}lling, F. Gerbier, A. Widera, O. Mandel, T. Gericke, and I. Bloch,
     Nature {\bf 434}, 481 (2005).
%%%Compressibility
\bibitem{in-situ_Mott:Nat}
     N. Gemelke, X. Zhang, C-L. Hung and C. Chin,
     Nature {\bf 460}, 995 (2009)
%Single site
\bibitem{singleSiteMPQ:Nat}
     J. F. Sherson, C. Weitenberg, M. Endres, M. Cheneau, I. Bloch, and S. Kuhr,
     Nature {\bf 467}, 68 (2010).
\bibitem{singleSiteHarvard:Sci}
     W. S. Bakr, A. Peng, M. E. Tai, R. Ma, J. Simon, J. I. Gillen, S. F\"{o}lling, L. Pollet, M. Greiner,
     Science {\bf 329}, 547 (2010).
%%%Shell Structure
%\bibitem{MRI_Folling:PRL}
%     S. F\"{o}lling, A. Widera, T. M\"{u}ller, F. Gerbier, and I. Bloch,
%     Phys. Rev. Lett. {\bf 97}, 060403 (2006).
\bibitem{RF_Gerbier:PRL}
     F. Gerbier, S. F\"{o}lling, A. Widera, O. Mandel, and I. Bloch,
     Phys. Rev. Lett. {\bf 96}, 090401 (2006).
\bibitem{Campbell}
     G. K. Campbell, J. Mun, M. Boyd, P. Medley, A. E. Leanhardt, L. G. Marcassa, D. E. Pritchard and W. Ketterle,
     Science {\bf 313}, 649 (2006).
%%%Bragg
%1D
\bibitem{2009:Bragg:Inguiscio}
     D. Cl\'{e}ment, N. Fabbri, L. Fallani, C. Fort, and M. Inguscio,
     Phys. Rev. Lett. {\bf 102}, 155301(2009).
%3D
\bibitem{2010:Sengstock:Bragg}
     P. T. Ernst, S. G\"{o}tze, J. S. Krauser, K. Pyka, D.-S. L\"{u}hmann, D. Pfannkuche, and K. Sengstock,
     Nature Phys. {\bf 6}, 56 (2010).
\bibitem{2010:Bragg3D:Heinzen}
	X. Du, S. Wan, E. Yesilada, C. Ryu, D. J. Heinzen, Z. Liang, and B. Wu,
     New J. Phys. {\bf 12}, 083025 (2010).

%%%Mott Gap
\bibitem{LatticeMod:PRL}
     M. J. Mark, E. Haller, K. Lauber, J. G. Danzl, A. J. Daley, and H.-C. N\"{a}gerl,
     Phys. Rev. Lett. {\bf 107}, 175301 (2011).
%%%%%%%%%%%%%%%%%%%%%%%%%%%%%%%%%%%%%%%%%%%%%%%%%%%%%%%%

\bibitem{Hazzard}
K. R. A. Hazzard and E. J. Mueller, Phys. Rev. A {\bf 76}, 063612 (2007).
\bibitem{Hazzard2010}
K. R. A. Hazzard and E. J. Mueller, Phys. Rev. A {\bf 81}, 033404 (2010).

\bibitem{Yamashita}
M. Yamashita and M.~W. Jack, Phys. Rev. A {\bf 79}, 023609-7 (2009).


\bibitem{Oktel}
M. \"O. Oktel, T. C. Killian, D. Kleppner, and L. S. Levitov
Phys. Rev. A, {\bf 65}, 033617 (2002)


%%Theory
%\bibitem{MFT:PRB}
%     B. Capogrosso-Sansone, N. V. Prokof�E�E�E�E�E�E�E�E�E�E�E�E�E�E�E�ev, and B. V. Svistunov,
%     Phys. Rev. B {\bf 45}, 3137 (1992).
%\bibitem{SBMethod:PRA}
%     D. B. M. Dickerscheid, D. van Oosten, P. J. H. Denteneer, and H. T. C. Stoof,
%     Phys. Rev. A {\bf 68}, 043623 (2003).
%\bibitem{QMC:PRA}
%     S. Wessel, F. Alet, M. Troyer, and G. G. Batrouni,
%     Phys. Rev. A {\bf 70}, 053615 (2004).
%\bibitem{DeMarco:PRA}
%     B. DeMarco, C. Lannert, S. Vishveshwara, and T.-C. Wei,
%     Phys. Rev. A {\bf 71}, 063601 (2005).
%\bibitem{QMCFiniteTemp:PRB}
%     B. Capogrosso-Sansone, N. V. Prokof�E�E�E�E�E�E�E�E�E�E�E�E�E�E�E�ev, and B. V. Svistunov,
%     Phys. Rev. B {\bf 75}, 134302 (2007).
%\bibitem{QMCTemp:NJP}
%     L. Pollet, C. Kollath, K. V. Houcke, and M. Troyer,
%     New J. Phys. {\bf 10}, 065001 (2008).
%\bibitem{Yamashita:PRA}
%     M. Yamashita and M. W. Jack,
%     Phys. Rev. A {\bf 79}, 023609 (2009).
%%%%%%%%%%%%%%%%%%%%%%%%%%%%%%%%%%%%%%%%%%%%%%%%%%%%%%%%
%%QCP
%\bibitem{Chin_QCP:NJP}
%     X. Zhang, C.-L. Hung, S.-K. Tung, N. Gemelke and C. Chin,
%     New. J. Phys. {\bf 13}, 045011 (2011).
%%%%%%%%%%%%%%%%%%%%%%%%%%%%%%%%%%%%%%%%%%%%%%%%%%%%%%%%
%%SF-NS
%\bibitem{Trotzky:NPhys}
%     S. Trotzky, L. Pollet, F. Gerbier, U. Schnorrberger, I. Bloch, N. V. Prokofev, B. Svistunov and M. Troyer,
%     Nat. Phys. {\bf 6}, 998 (2010).
%%%%%%%%%%%%%%%%%%%%%%%%%%%%%%%%%%%%%%%%%%%%%%%%%%%%%%%%
%%3P2
\bibitem{NJP:507}
     A. Yamaguchi, S. Uetake, S. Kato, H. Ito, and Y. Takahashi,
     New J. Phys. {\bf 12}, 103001 (2010).
%\bibitem{3P2col:PRL}
%     A. Yamaguchi, S. Uetake, D. Hashimoto, J. M. Doyle, and Y. Takahashi,
%     Phys. Rev. Lett {\bf 101}, 233002 (2008).
\bibitem{3P2MRI:App}
     S. Kato, K. Shibata, R. Yamamoto, Y. Yoshikawa, and Y. Takahashi,
		Appl. Phys. B {\bf 108}, 31 (2012).
\bibitem{YamashitKato}
     M. Yamashita, and S. Kato, A. Yamaguchi, S. Sugawa, T. Fukuhara, S. Uetake, and Y. Takahashi,
		Phys. Rev. A {\bf 87}, 041604 (2013).

%%%%%%%%%%%%%%%%%%%%%%%%%%%%%%%%%%%%%%%%%%%%%%%%%%%%%%%%
%%Yb QDs
%\bibitem{2007:Fukuhara:FD}
%	Takeshi Fukuhara, Yosuke Takasu, Mitsutaka Kumakura, and Yoshiro Takahashi,
%	Phys. Rev. Lett. {\bf 98}, 030401(2007).
%
%\bibitem{2009:Fukuhara:Mix}
%	Takeshi Fukuhara, Seiji Sugawa, Yosuke Takasu, and Yoshiro Takahashi,
%	Phys. Rev. A {\bf 79}, 021601(2009).

%\bibitem{2010:Taie:SU}%
%	Shintaro Taie, Yosuke Takasu, Seiji Sugawa, Rekishu Yamazaki, Takuya Tsujimoto, Ryo Murakami, and Yoshiro Takahashi,
%	Phys. Rev. Lett. {\bf 105}, 190401(2010).

\bibitem{2011:Sugawa:DualMott}
	S. Sugawa, K.e Inaba, S. Taie, R. Yamazaki, M. Yamashita and Y. Takahashi,
	Nat. Phys. {\bf 7}, 642(2011).
%%OT
%\bibitem{OTMIT:PRL}
%     T. L. Gustavson,A. P. Chikkatur, A. E. Leanhardt, A. G\"{o}rlitz, S. Gupta, D. E. Pritchard and W. Ketterle,
%     Phys. Rev. Lett. {\bf 88}, 020401 (2001).

\bibitem{ExpPaper}
S. Kato, private communication.

%Hubbard
\bibitem{BH:PRB}
     M. P. A. Fisher, P. B. Weichman, G. Grinstein, and D. Fisher,
     Phys. Rev. B {\bf 40}, 546 (1989).
\bibitem{Jaksch:PRL}
     D. Jaksch, C. Bruder, J. I. Cirac, C. W. Gardiner, and P. Zoller,
     Phys. Rev. Lett. {\bf 81}, 3108 (1998).


\bibitem{BandMapping}
M. Greiner, I. Bloch, O. Mandel, T. W. H\"{a}nsch, and T. Esslinger, Phys. Rev. Lett. {\bf 87}, 160405 (2001)


%\bibitem{Georeges}
%A. Georges, G. Kotliar, W. Krauth and M.~J. Rozenberg, Rev. Mod. Phys. {\bf 68}, 13 (1996).

%\bibitem{Potthoff}
%M. Potthoff, Eur. Phys. J. B {\bf 32}, 429-436 (2003).


\end{references}
\end{document}